\documentclass[11pt,twocolumn]{article}
\usepackage{times}
\usepackage{helvet} 
\usepackage{courier}

\usepackage{amsmath,amsthm}     
\usepackage{graphicx}     
\usepackage{hyperref} 
\usepackage{url}
\usepackage{amsfonts} 

\usepackage[numbers]{natbib} 

\usepackage{geometry}
\theoremstyle{theorem}

\theoremstyle{definition}

\newtheorem*{remark}{Remark}

\title{A Kinematic and Geometric Analysis of Trochoidal Waves}
\author{Andrew D Irving\\Independent Researcher\\
{Email: a${\_}$irving@btinternet.com} \and Ebrahim L Patel\\
University of Greenwich\\
{Email: e.patel@greenwich.ac.uk}}

\date{}

\begin{document}
\maketitle







\section*{Abstract}
To study the geometry of 
Gerstner's 
water wave 
model, 
we analyse the velocity of his fluid particles 
in a 
reference 
frame 
that moves 
with the wave. 
Gerstner wave profiles are 
cycloidal,
curtate (flattened)  
trochoids, 
or 
prolate (extended) trochoids.
We  
derive 
both 
the height of each 
profile's 
characterising point 
(cusp, 
inflection, 
or  
self-intersection), as well as a condition 
under which the arc lengths of prolate and curtate 
profiles coincide over a single wave cycle. 
We conclude
with a discussion of how Galilean 
transformations affect particle acceleration and the 
geometry of their trajectories.

\section{Introduction}
\begin{figure}[h]
\renewcommand\thefigure{1} 
\begin{center}
\includegraphics[width=0.25\textwidth]{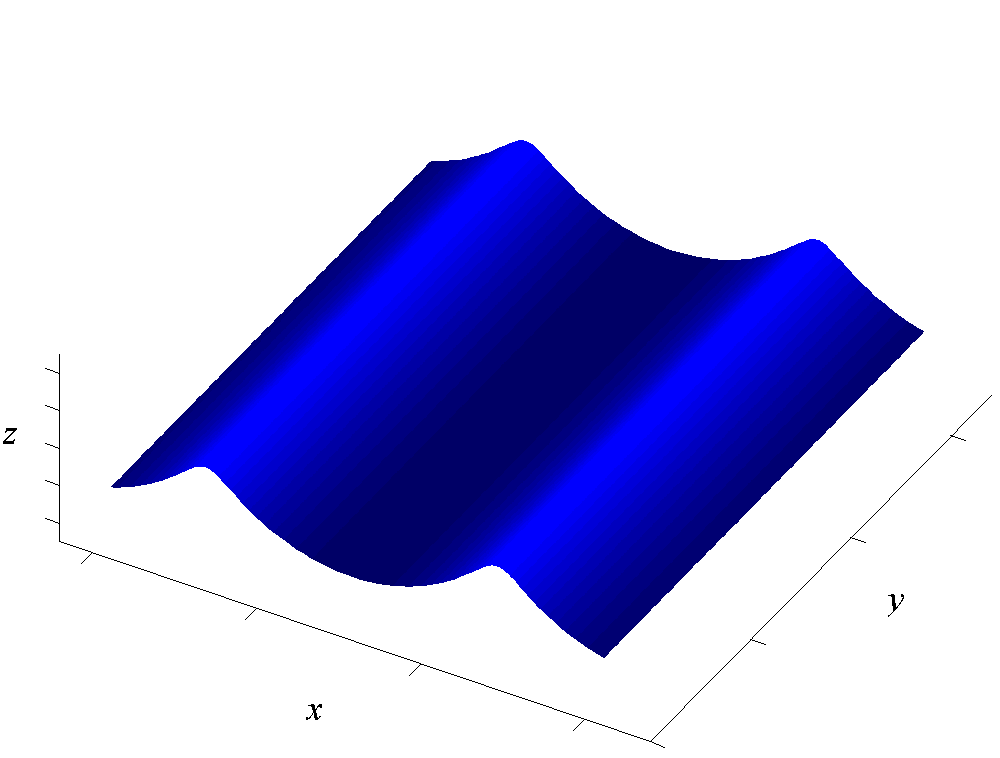}
\includegraphics[width=0.25\textwidth]{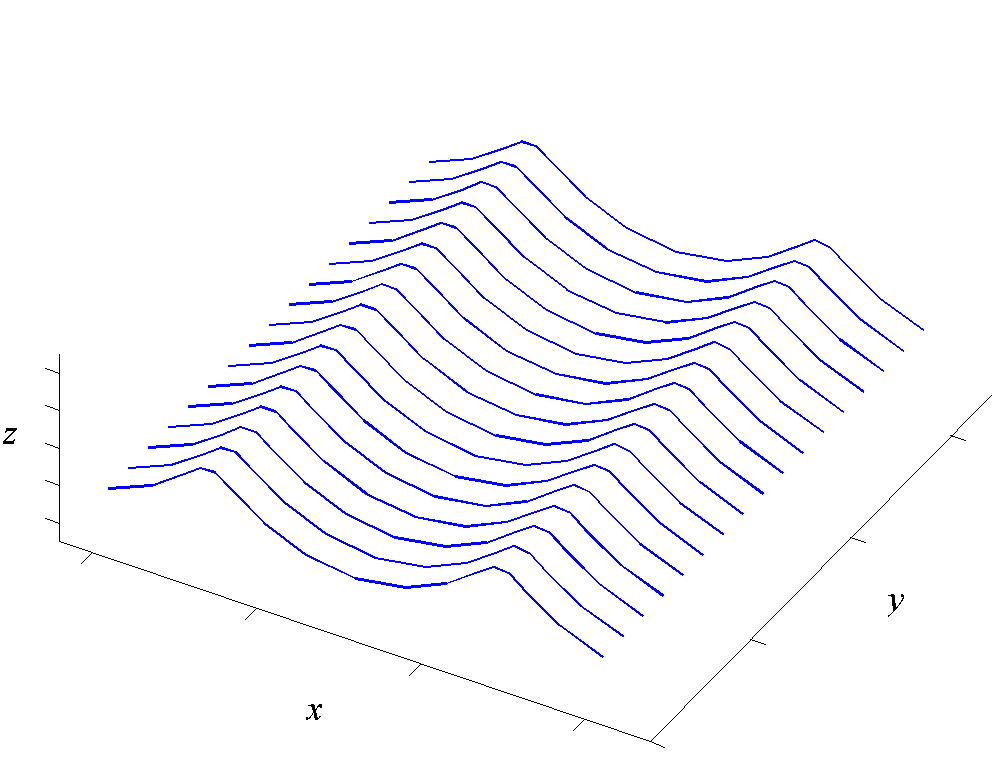}
\caption{Simulation of~\protect\cite{Flick}.
Surface 
governed by~(\ref{xandy})
and $y = y$ at time $t$ (top).  
Cross sections defined by 
displaying 
surface 
only for integer values of $y$ (bottom). 
Gerstner's model 
(see~(\ref{xandy}))
describes the $y = 0$ cross section.} 
\label{gerstner2D}
\end{center}
\end{figure}

\noindent Among the computer graphics community, 
extensions to 
a 
celebrated 
mathematical model have found favour 
in the rendering of rough seas (e.g. see 
Figure~\ref{gerstner2D}).  
%
%
%
%
Czech 
polymath 
Franti\v{s}ek
Gerstner 
pioneered 
the original model 
more than two centuries ago~\cite{Gerstner}. 
The model 
would be 
rediscovered independently 
some sixty years later 
in the study 
of 
rolling 
ships~\cite{Henry}.  
Like Gerstner, 
English Naval Architect William Froude 
and 
Scottish 
Engineer William Rankine 
ambitiously 
follow a Lagrangian approach: 
specifying 
the evolution of 
each 
fluid particle.
Yet the model is surprisingly simple and 
produces visually realistic surface 
waves~\cite{Froude,Rankine}. 

A wind blowing across the ocean 
surface causes most such waves:  
the moving air 
displaces
the water from 
a level surface 
equilibrium. 
%
Surface tension and gravity may then combine as 
restorative forces.
Thus, the fluid moves back and 
forth.  When this occurs periodically, the movement 
gives rise to waves.  
The effect of 
surface tension 
can be 
assumed negligible 
if 
the 
wavelength 
exceeds 
two 
centimetres.  
Therefore, 
it is 
gravity 
which 
dominates 
the fluid 
dynamics 
to give 
these longer waves 
the name of 
gravity waves~\cite{Thurman}.  

Newton's law of inertia 
predicts 
a 
moving body 
to 
follow 
a 
linear path,  
unless 
that body is 
acted upon by 
some 
force(s)~\cite{Ganot}. 
Such forces are understood to act upon 
the 
water molecules that form gravity waves.     
Hence, Gerstner 
modelled 
fluid particles 
to 
deviate from linear trajectories, following curved 
orbits instead. 
As a result of this deviation,  
the wave profiles 
that Gerstner's 
fluid particles 
generate are not 
sinusoidal, but trochoidal.

In the rest of this paper, we highlight the role of particle velocity in the differential geometry of Gerstner waves: their curvature, changes in concavity, singularities, and arc length.

\section{Fluid particles: Cross section of gravity waves}
\begin{figure}[h]
\begin{center}
\includegraphics[width=0.5\textwidth]{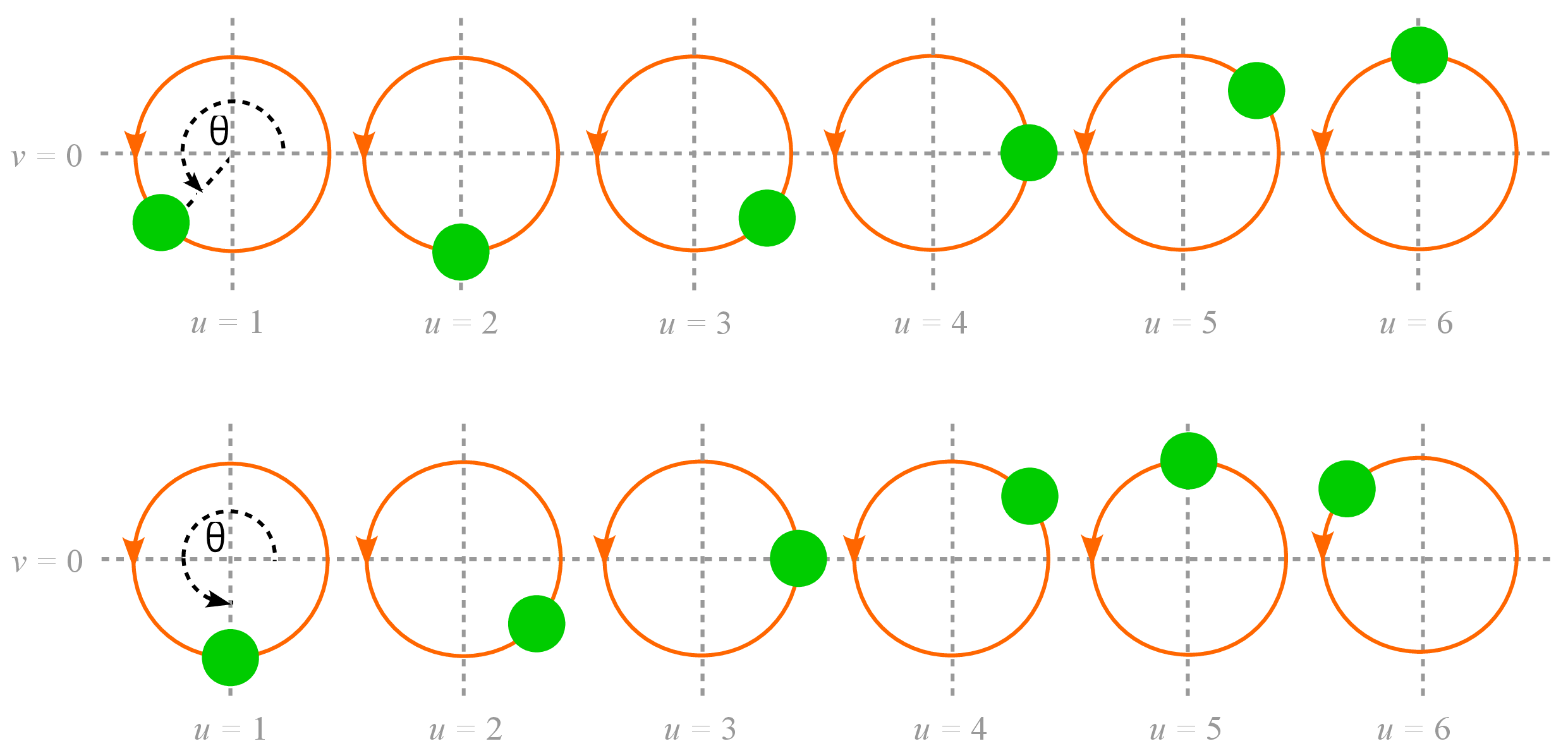}
%
%
%
\caption{Sample of 
surface 
particles 
(green) at time $t$ (top)
and
an eighth of
a cycle
later (bottom).
In the $xz-$plane, 
a 
particle 
subtends an angle $\theta$ at the centre
of a fixed
orbit (orange).
Here, we employ~(\ref{xandy})
using 
$\omega > 0$, 
$r < 1 / 2$, and only 
integer values of $u$.}
\label{thetadiag}
\end{center}
\end{figure}
Let us 
begin by introducing 
Gerstner's model.  
A snapshot of an ocean wave may show it to 
be a regular pattern of peaks and troughs~\cite{Thomas}.
None of these peaks or troughs are single points.  
Instead, 
Gerstner treats 
each peak or trough as a straight line.
All such lines are parallel  
(e.g. see 
Figure~\ref{gerstner2D}). 
In formal language, Gerstner assumes the ocean surface  
to be a 
{\em two-dimensional wave}.  

Figure~\ref{gerstner2D} illustrates the great advantage 
of such a model.  Any cross section 
of the wave which is 
perpendicular to a linear peak or trough is equivalent.
%
%
%
%
We can therefore study the behaviour 
of a two-dimensional wave through any 
such 
cross section.

Gerstner 
pictures the  
two-dimensional wave 
in $xyz-$space 
such that one of his
equivalent 
cross sections lies entirely in the $xz-$plane 
(as in Figure~\ref{gerstner2D}).
For any 
$u$ and $v \le 0$, Gerstner proposes there to be 
a 
fluid particle 
of coordinates,
%
%
\begin{equation}
\label{xandy}
x = r \cos \theta + u 
~~~\mbox{ , }~~~
z = r \sin \theta + v
\end{equation}
within that cross section. 


The  
particle
subtends an angle of 
$\theta$ 
at 
a 
fixed 
centre. 
That centre has coordinates $( u , v )$ in 
the $x z-$plane 
(e.g. see Figure~\ref{thetadiag}).
Angle 
$\theta = \omega t + G u$ 
increases by $\omega$ radians with every unit of 
time $t$ that passes 
(making $\omega$ the 
particle's 
angular frequency 
and $G u$ the initial $\theta$, where $\omega$ and 
$G$ are non-zero parameters).

Hence, this 
particle 
traces a 
counterclockwise 
circular 
orbit 
around $( u , v )$ over time when $\omega > 0$. 
Depending on the altitude of its centre, 
this orbit has 
radius $r = e^{m v} / m$,  
where $m > 0$ is a constant~\cite{Henry}.  





\section{Composition}
\begin{figure}[h]
\begin{center}
\includegraphics[width=0.3\textwidth]{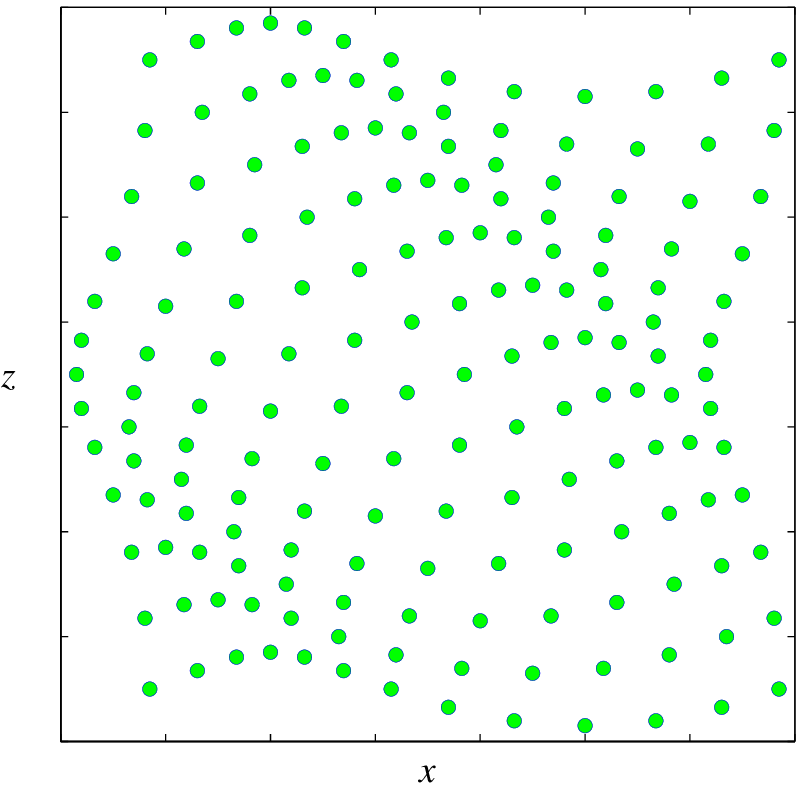}
\includegraphics[width=0.3\textwidth]{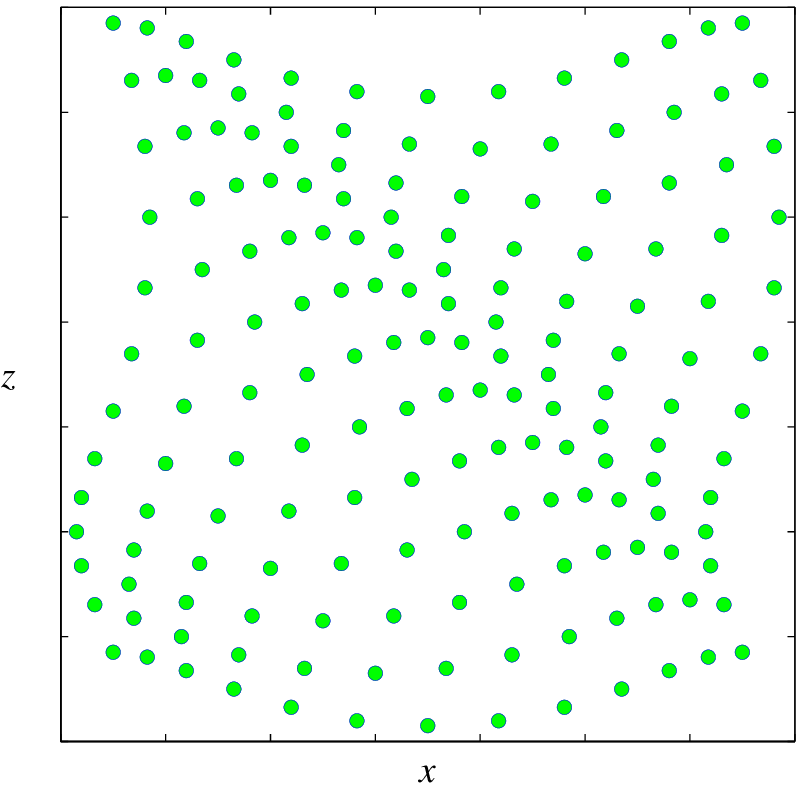}
\caption{Our simulation of Emma Phillips' animation 
in~\protect\cite{Phillips}.  One hundred and sixty nine 
green dots 
at time $t$ (top) and
later instant 
(bottom).
Collectively, dots generate
waves which travel in directions of 
decreasing $x$ and decreasing $z$.}
\label{phillips}
\end{center}
\end{figure}
\noindent
Even outside the computer graphics community,  
Gerstner's 
fluid particles 
have 
made 
an 
impression,
thanks 
to
an 
animation 
that has been widely 
imitated 
online 
(e.g. see Figure~\ref{phillips}).  
Here, 
we 
first 
prove Phillips' animation 
to be an adaptation of Gerstner's 
work.  
As such, 
the 
animation is instructive as 
a first introduction to the structure of water 
waves.

In~\cite{Phillips}, 
Phillips' dots 
perform 
uniform circular motion
such that 
all dots share a single fixed speed.  
Perhaps unexpectedly then,
neighbouring dots draw closer 
over one time period before 
pulling apart 
over the next 
(see Figure~\ref{phillips}).  
As we shall prove, 
this is no optical 
illusion.  

Behind Emma Phillips' eye-catching 
compositions are 169 dots 
tracing circular paths of radius 
$1 / m$.  Given an integer 
$u \in [ -6 , 6 ]$ and an integer 
$v \in [ -6 , 6 ]$, Phillips 
renders a dot which subtends an angle 
of 
$\omega t + G u + G v$ 
at $( u , v )$ in the $x z-$plane~\cite{Danovich}.  
Hence, Gertsner's 
model governs 
those 
dots 
defined 
by Phillips' $u$ values and $v = 0$. 

\begin{figure}[h]
\begin{center}
\includegraphics[width=0.3\textwidth]{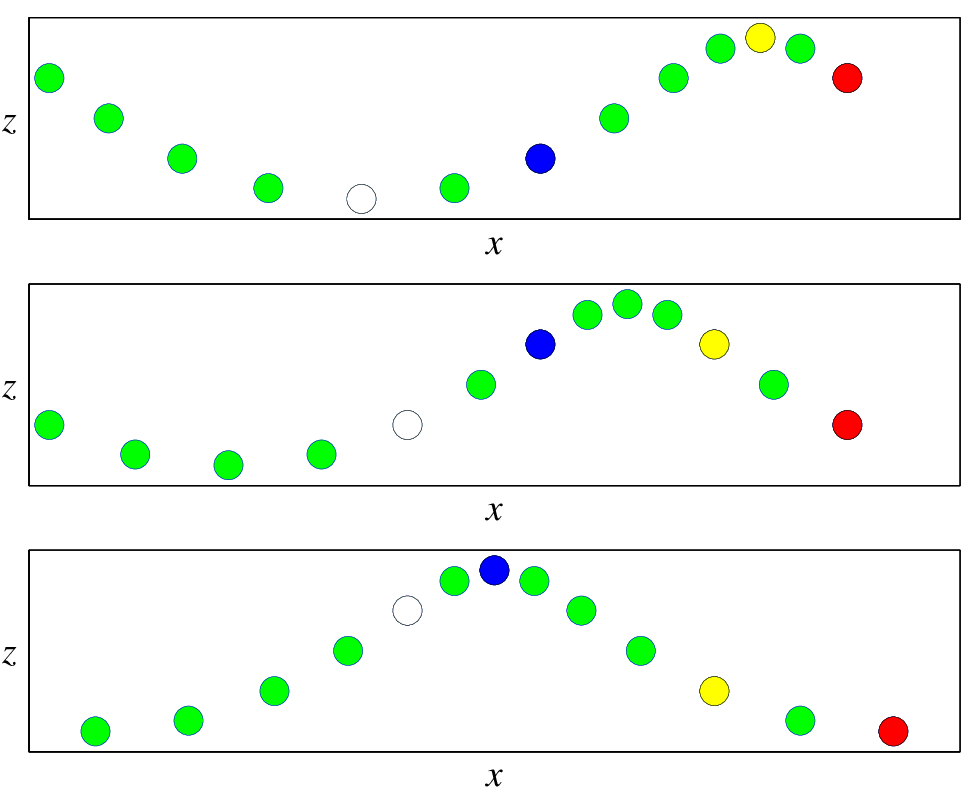}
\includegraphics[width=0.3\textwidth]{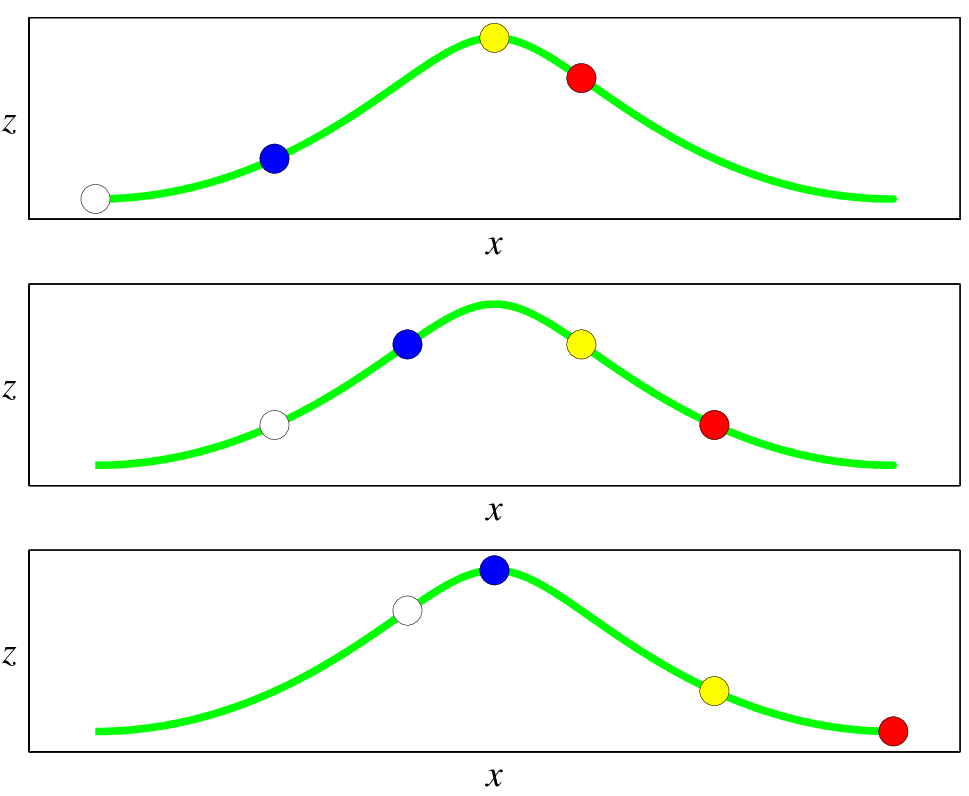}
\caption{Sample of 
Phillips' dots 
at
three moments
spanning
a
third of a
period 
($t$ increases top to bottom).
In a fixed frame (upper images), 
dots trace circular paths with velocity \textbf{D} 
and wave 
moves leftward. 
In a wave cycle's frame (lower images),
the 
same 
dots 
have velocity \textbf{R}, 
moving rightward as they 
trace a (green) trochoidal cycle  
($r G < 1$ here).}
\label{frame}
\end{center}
\end{figure}

Such a sample is enough to show us 
how a Gerstner wave propagates.
As Figures~\ref{thetadiag} and~\ref{frame} illustrate, 
the peaks and troughs of this wave are just points of 
constant $\theta$ which advance from one dot to 
another~\cite{Raymond}.
In Figure~\ref{thetadiag} for instance, the 
peak is a point where $\theta = 5 \pi / 2$ while 
the trough is a point where $\theta = 3 \pi / 2$.

%

Let $P$ be 
the point 
on Phillips' 
Gerstner wave where $\theta$ is a constant, $c$.
Let $D$ 
be 
the 
dot which orbits $( b , 0 )$ 
in the $x z-$plane.
Thus, 
$P$ 
coincides with 
$D$
if and only if 
$D$ subtends an angle of 
$c$ at $( b , 0 )$.
This occurs 
when $D$ has coordinates,
\begin{eqnarray}
\label{frontX}
x &=&
r \cos c + b
\\
\label{frontY}
z &=& r \sin c
\end{eqnarray}
noting we have simply input $\theta = c$ 
and $( u , v ) = ( b , 0 )$ into~(\ref{xandy}).
This locates $P$ for any given $b$. 

However, we wish to be able to locate $P$ 
more generally.
Using 
$\theta = c$, we can 
replace $b$ in~(\ref{frontX}) with 
$( c - \omega t ) / G$. 
The resulting~(\ref{frontX}) and~(\ref{frontY}) 
gives us,
%
%
\begin{equation}
\label{line}
G z + {G} x =
{G} r \cos c + G r \sin c
+ {c - \omega t}
\end{equation}
which 
locates 
$P$ 
at time $t$.
Therefore, 
the 
time-derivative of~(\ref{line}),
\begin{equation}
\label{pfrontvel}
\frac{\mathrm{d} z}{\mathrm{d} t} + 
\frac{\mathrm{d} x}{\mathrm{d} t} = 
- \frac{\omega}{G}
\end{equation}
governs the 
velocity 
of $P$. 
Unlike~(\ref{line}),~(\ref{pfrontvel}) does not 
depend on $c$.
%
This 
independence of $c$ 
tells us 
that~(\ref{pfrontvel}) 
governs 
not just $P$, but any 
peak or trough 
in Figures~\ref{thetadiag} or~\ref{frame}.


Those 
peaks or troughs 
maintain fixed heights.   
As 
such, 
they are 
governed by 
${\mathrm{d} z} / {\mathrm{d} t} = 0$.   
By substituting 
${\mathrm{d} z} / {\mathrm{d} t} 
= 0$ into~(\ref{pfrontvel}), we find 
their 
velocity
\textbf{P}
to be  
${\mathrm{d} x} / {\mathrm{d} t} = 
- \omega / G$. 
According to~(\ref{xandy}) then, 
$D$ 
exhibits motion in both the 
transverse and longitudinal 
directions to a passing wave.  
To this extent, 
$D$ emulates a 
water molecule~\cite{Considine}.

Further reinforcing this comparison, 
the derivatives of~(\ref{xandy})
show 
the velocity of 
$D$ to be,
\begin{equation}
\label{instvel}
\textbf{D} = 
- r \omega ( \sin \theta ) \, \textbf{i}
+ r \omega ( \cos \theta ) \, \textbf{k}
= 
- \omega z \, \textbf{i}
+ \omega ( x - b ) \, \textbf{k}
\end{equation}
%
%
%
at time $t$, 
where
$\textbf{i}$
and $\textbf{k}$
are unit vectors in the horizontal 
direction 
and
the
vertical direction respectively.
Dot $D$ thus behaves 
like a water molecule:  
moving
with
an
oncoming wave at its 
peaks 
(where $x = b$ and $z = r$ in~(\ref{instvel}))
 and against
the wave at its 
troughs
(where $x = b$ and $z = -r$ 
in~(\ref{instvel}))~\cite{King}.

Although 
fluid 
moves 
forward 
with a water 
wave at its peaks, 
Stokes predicted in 1880 that the wave would 
become unstable 
if 
fluid 
were to 
move forward faster than the wave itself~\cite{TMD}.
Or if you prefer, the wave will {\em break} 
if 
fluid 
moves forward relative to the wave.

Fluid particles 
are 
thus 
typically 
modelled
to
flow
backwards
relative
to
the
forward-moving
surface wave~\cite{Toda} (as in  
Figure~\ref{frame}).  
With respect 
to the 
wave 
that moves forward with velocity \textbf{P},
$D$ does 
not trace a circular path 
with velocity
\textbf{D}. 
Instead, 
the laws of Galilean relativity dictate that 
$D$ 
must move 
with a velocity,
%
\begin{equation}
\label{relvel}
\textbf{D} - \textbf{P} =
\omega
\left(
\frac{1}{G}
- r \sin \theta \right) \, \textbf{i}
+ \omega ( r \cos \theta ) \, \textbf{k}
\end{equation}
%
%
%
%
that 
we 
shall 
denote \textbf{R}~\cite{Greenhill,Lumley}. 
Hence, 
fluid particles 
flow 
backwards 
relative to 
their 
wave 
when 
the horizontal component of \textbf{R} 
has the opposite sign 
to that of 
\textbf{P}. 
Consequently, 
water 
waves 
remain stable 
when 
$r G < 1$. 

\begin{remark}
The 
horizontal components of \textbf{R} and 
\textbf{P} have opposite signs 
if
$( 1 / G )
- 
r \sin \theta
> 0$. 
Given the left hand side of
this inequality has minimal value
$( 1 / G )
- r$,
the 
inequality holds 
for all 
$\theta$ 
when
$r G < 1$. 
\end{remark}



Equation~(\ref{relvel})
tells us how 
water 
behaves 
in the eyes of sailors who
match the wave velocity.
According to~(\ref{relvel}), 
such 
sailors 
will
perceive  
fluid particles 
to 
trace 
an (upside-down) 
{\em trochoid}~\cite{Nelson}.
This  
trochoidal shape 
can be  
seen in Nature, 
on the surface of ocean waves~\cite{Thurman}.  
This is no coincidence.  
As Figure~\ref{frame} illustrates, 
sailors 
perceive 
fluid 
moving along 
the 
surface 
of the 
ocean 
wave.


By studying 
particles 
from this 
perspective, 
we 
learn 
about the waves 
they trace.
When $r < 1 / G$ in~(\ref{relvel}),
the horizontal component of \textbf{R} 
strictly increases in 
size from peak to trough.  
Fluid particles 
therefore 
travel 
further backward in tracing
the lower half of a 
stable 
trochoidal 
wave
cycle
than they
do
in
tracing
its
upper half. 
Relative
velocity \textbf{R} thus
predicts 
the
lower
half of
any
stable 
trochoidal
wave cycle 
to be 
wider
than its
upper
half~\cite{Hyper}.
Whether the wave is 
stable or unstable, 
\begin{equation}
\label{relspeed}
\left| \textbf{R} \right| = 
\left| \omega \right| 
\sqrt{ \frac{1}{G^2} - \frac{2}{G}
r \sin \theta
+ r^2 }
\end{equation}
defines the speed that 
$D$ exhibits 
in tracing 
the trochoidal wave profile.
With respect to
any
such
cycle,~(\ref{relspeed}) shows that
the speed of
the
fluid particles 
is negatively correlated to their height.

As 
two such 
particles 
rise towards a 
trochoidal 
peak,~(\ref{relspeed})
tells us that 
the 
leading (higher) 
particles 
is 
the 
slower at 
each 
instant.
Over this period, 
the trochoidal arc 
length 
between 
the two  
therefore 
decreases.
As 
two such 
particles 
fall 
towards a trough,~(\ref{relspeed})
tells us that 
the 
leading (now lower) 
particle 
is 
the 
faster 
at each instant.
Over this period, 
the trochoidal arc 
length 
between 
the two 
therefore 
increases.


The composition of~\cite{Phillips} reflects this. 
When dot 
$D$
subtends an angle of  
$\theta \equiv \frac{\pi}{2} \bmod 2\pi$ 
at $( x , z ) = ( b , 0 )$, 
$D$ is closer to 
dots which orbit 
$( x , z ) = 
( b \pm 1 , 0 )$
than 
it is 
when 
$D$ 
subtends an angle of  
$\theta \equiv \frac{3 \pi}{2} \bmod 2\pi$
at $( x , z ) = ( b , 0 )$
(see Figure~\ref{thetadiag}). 

\section{Curvature}
\begin{figure}[h]
\begin{center}
\includegraphics[width=0.48\textwidth]{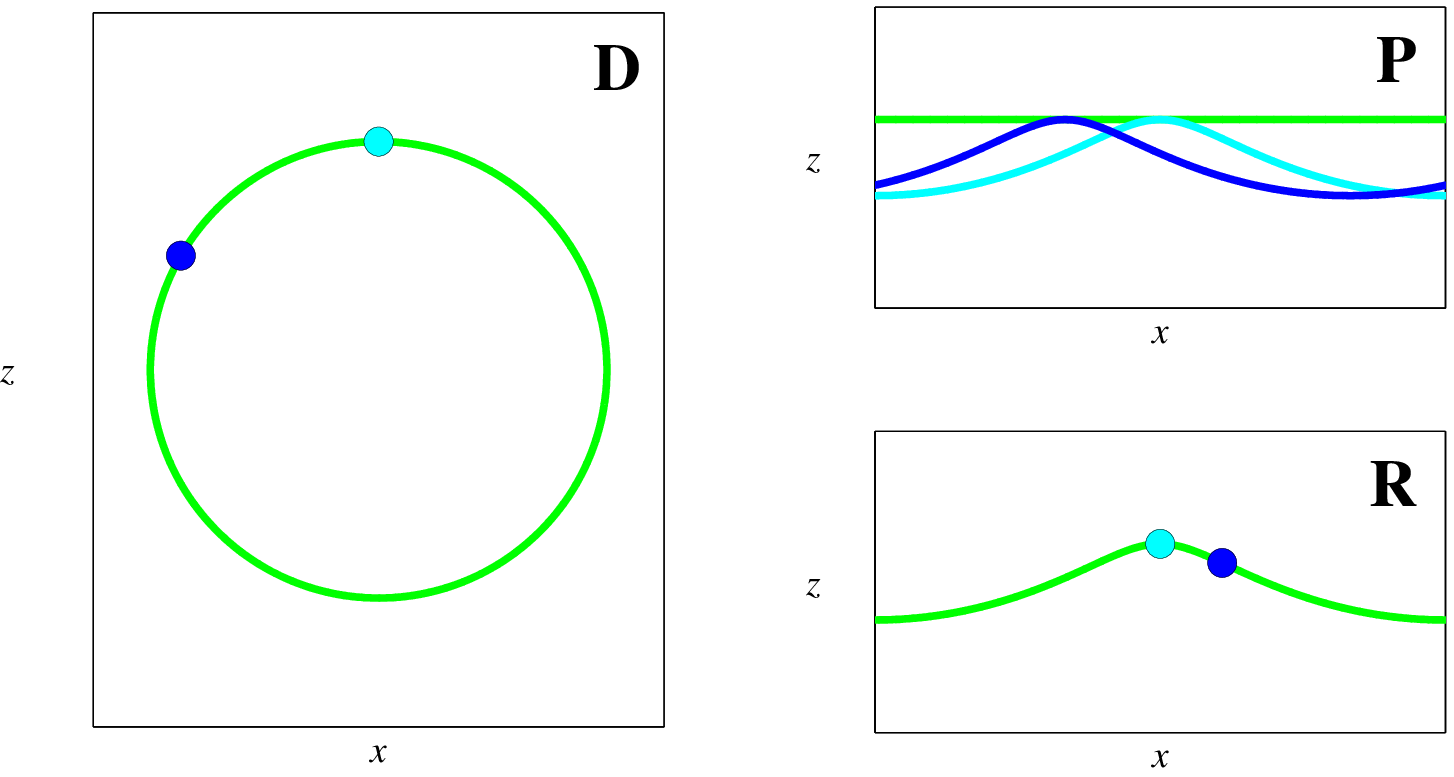}
\caption{Trajectory and associated 
velocity of fluid particle
illustrated: in fixed 
frame (left), and in frame of the 
wave  
(bottom right) 
($\textbf{R} = \textbf{D} - \textbf{P}$, 
where \textbf{P} defines 
the (phase) velocity of the wave). 
Paths 
traced by 
particles and wave peak marked in green.
Positions of particles
and peaks marked 
at time $t$ (cyan) and a fraction of a cycle later (blue).} 
\label{3vel}
\end{center}
\end{figure}
\noindent
In the previous section, 
we 
began to see 
how 
\textbf{R} - 
the velocity of water in a sailor's 
frame of reference - 
can 
communicate the structure of water 
waves.
To 
explore 
this structure 
more deeply, 
we 
shall 
introduce
the mathematics
that 
most 
readily decipher   
acceleration.

As Figure~\ref{3vel} highlights,
Phillips' 
fluid particle
$D$ traces 
the trochoidal 
profile 
of a Gerstner wave 
with
acceleration 
${\mathrm{d} \textbf{R}} / {\mathrm{d} t}$.  
Using~(\ref{relvel}),
%
%
%
%
%
%
%
%
%
%
%
%
%
\begin{equation}
\label{Sderiv}
\frac{\mathrm{d} \textbf{R}}{\mathrm{d} t}
=
\frac{\mathrm{d} \textbf{D}}{\mathrm{d} t}
-
\frac{\mathrm{d} \textbf{P}}{\mathrm{d} t}
=
- r \omega^2 ( \cos \theta ) \, \textbf{i}
- r \omega^2 ( \sin \theta ) \, \textbf{k}.
\end{equation}
%
%
%
Given
${\mathrm{d} \textbf{P}} / {\mathrm{d} t} = 0$,
equation~(\ref{Sderiv})
says 
what at
first seems counterintuitive:
${\mathrm{d} \textbf{R}} / {\mathrm{d} t}
=
{\mathrm{d} \textbf{D}} / {\mathrm{d} t}$.
Hence,
a
body
that
alternately
quickens and slows
along
a
trochoidal
path
(see~(\ref{relspeed}))
changes velocity at the same rate as
a
body
which
travels at constant speed
along
a
circular
path
(see~(\ref{instvel})).


This is
no paradox. 
When 
a body 
traces a curved path 
with 
variable speed 
$| \textbf{v} |$,
the direction of 
motion changes at one rate while the speed 
changes at another rate.
%
%
%
As in 
equation~(\ref{Sderiv}), a horizontal 
component can be apportioned 
fractions of 
both rates.  So too a vertical component.  
Acceleration can thus prove difficult to 
interpret.
%
Hence,
a
preferred decomposition
of acceleration
is,
\begin{equation}
\label{accelformula}
%
%
\frac{\mathrm{d} \textbf{v}}{\mathrm{d} t} =
\frac{\mathrm{d}}{\mathrm{d} t}
\left( | \textbf{v} | \frac{\textbf{v}}{| \textbf{v} |} \right) =
\frac{\mathrm{d}}{\mathrm{d} t}
\left( | \textbf{v} | \, \textbf{T} \right)
=
\frac{\mathrm{d} | \textbf{v} |}{\mathrm{d} t}\,
\textbf{T}
+
| \textbf{v} |
\frac{\mathrm{d} \textbf{T}}{\mathrm{d} t}.
\end{equation}
Equation~(\ref{accelformula}) 
separates the two rates 
at the heart of 
acceleration: 
${\mathrm{d} | \textbf{v} |} / {\mathrm{d} t}$ 
and 
${\mathrm{d} \textbf{T}} / {\mathrm{d} t}$.
While 
the former equals the rate of change in 
{\em speed}, the significance of the latter is 
perhaps less obvious.
In fact, 
${\mathrm{d} \textbf{T}} / {\mathrm{d} t}$
is the rate of change in the {\em direction 
of motion}.
Let us explain. 

\begin{figure}[h]
\begin{center}
\includegraphics[width=0.48\textwidth]{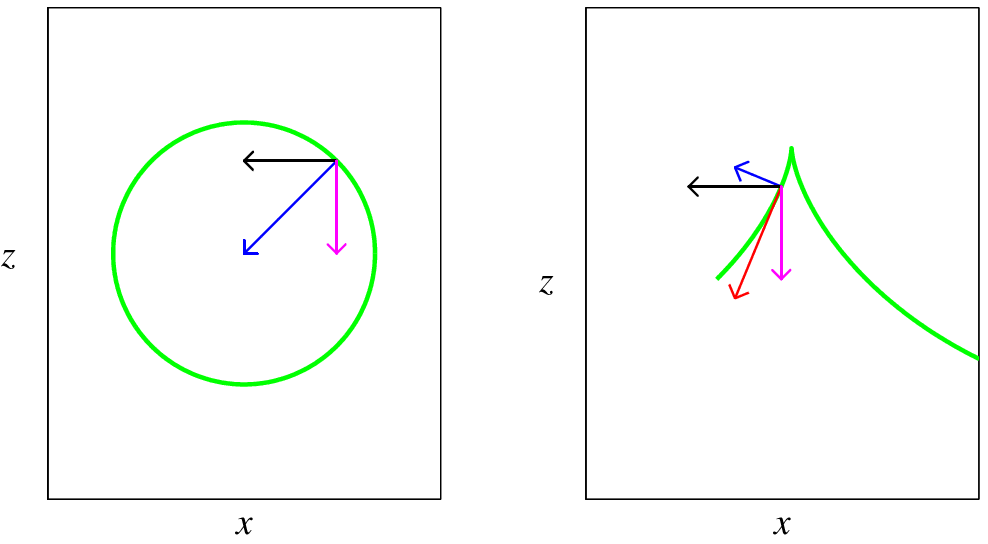}
\caption{Components of acceleration 
highlighted 
for 
paths 
(green) traced by 
particles 
of velocity \textbf{D} (left) and \textbf{R} (right).   
At 
points  
defined by 
$\theta = 2 \pi / 8$, 
the 
horizontal components (black) 
are equal and vertical components (pink) 
are equal.
However, tangential components (red) 
are not equal 
and 
normal 
components (blue) 
are not equal.}
\label{accelcomp}
\end{center}
\end{figure}

As a unit vector, 
$\textbf{T} = \textbf{v} / | \textbf{v} |$ 
cannot stretch or shrink over time: 
turning is all that $\textbf{T}$ can do.  
Therefore, 
$\mathrm{d} \textbf{T} / 
{\mathrm{d} t}$ - the rate of change in 
$\textbf{T}$ over time - 
is simply the rate at which 
$\textbf{T}$ 
turns.
Given $| \textbf{v} |$ 
is just a scalar, the 
direction of 
$\textbf{T} = \textbf{v} / | \textbf{v} |$ 
must equal the direction of $\textbf{v}$.  
Hence, 
\textbf{v} and \textbf{T} turn 
at the same 
rate.  
Consequently, 
${\mathrm{d} \textbf{T}} / {\mathrm{d} t}$ 
 is the rate 
of change in the direction of 
velocity 
\textbf{v}. 
By separating 
this  
rate 
from the rate 
at which vector 
$\textbf{v}$
stretches, 
equation~(\ref{accelformula}) can 
distinguish 
${\mathrm{d} \textbf{D}} / {\mathrm{d} t}$ 
and
${\mathrm{d} \textbf{R}} / {\mathrm{d} t}$. 
As Figure~\ref{accelcomp} illustrates, 
stretching 
rate 
${\mathrm{d} | \textbf{v} |} / 
{\mathrm{d} t} = 0$
when $\textbf{v} = \textbf{D}$:
the velocity vector 
turns 
without stretching
as a body
performs 
uniform circular motion.
In
contrast,
${\mathrm{d} | \textbf{v} |} / 
{\mathrm{d} t} \ne 0$
when $\textbf{v} = \textbf{R}$:
the velocity vector 
elongates and 
shortens as a body 
traces a trochoidal 
profile.


%
%

At first,~(\ref{accelformula})  
seems otherwise 
daunting 
to compute.  
Fortunately, 
we can extract value without 
computing each term.
We exploit a simple fact: \textbf{T} 
is 
parallel to itself.  
It follows that,
$\textbf{T} \cdot \textbf{T} = 
| \textbf{T} | | \textbf{T} |$.  
%
%
Given \textbf{T} is a unit vector, 
this means 
$\textbf{T} \cdot \textbf{T} = 
1$.  By differentiating both sides, 
we find that,
\begin{equation}
\label{perp}
\textbf{T} 
\cdot
\frac{\mathrm{d} \textbf{T}}{\mathrm{d} t} 
+ \frac{\mathrm{d} \textbf{T}}{\mathrm{d} t} 
\cdot
\textbf{T} 
=
\frac{\mathrm{d}}{\mathrm{d} t}
\left( 1 \right) 
\Longrightarrow
2 \left( \textbf{T} 
\cdot
\frac{\mathrm{d} \textbf{T}}{\mathrm{d} t} 
\right)
= 0 
\end{equation} 
 %
 %
 %
i.e. 
the dot product of \textbf{T} 
and 
${\mathrm{d} \textbf{T}} / {\mathrm{d} t}$ 
is zero: they are {\em perpendicular} vectors.

This is the final  
property we needed to 
take full advantage of~(\ref{accelformula}).
As
$\textbf{T} = \textbf{v} / | \textbf{v} |$
told us, 
\textbf{T} points in the 
direction of motion.  
Hence, \textbf{T} is {\em tangent} to a 
path traced by a body of velocity \textbf{v}.    
As~(\ref{perp}) just told us,
${\mathrm{d} \textbf{T}} / {\mathrm{d} t}$ 
is 
orthogonal 
to \textbf{T}.  Thus  
${\mathrm{d} \textbf{T}} / {\mathrm{d} t}$ 
is
{\em normal} 
to that same path.
Equation~(\ref{accelformula}) resolves 
${\mathrm{d} \textbf{v}} / {\mathrm{d} t}$
into 
$a_T \textbf{T}$ and 
$a_N 
{\mathrm{d} \textbf{T}} / {\mathrm{d} t}$: 
components we now know to be 
orthogonal 
to one another.

It follows that the 
acceleration 
${\mathrm{d} \textbf{v}} / {\mathrm{d} t}$ 
is the hypotenuse of a right-angled 
triangle
with 
legs 
$a_T \textbf{T}$ and 
$a_N 
{\mathrm{d} \textbf{T}} / {\mathrm{d} t}$.
By the Pythagorean 
theorem then,
%
%
\begin{equation*}
\left| \frac{\mathrm{d} 
\textbf{v}}{\mathrm{d} t} \right|^2
=
\left| \,{a_T}
\textbf{T} \right|^2 +
\left|
\,{a_N}
\frac{\mathrm{d} \textbf{T}}{\mathrm{d} t}
\right|^2 
\end{equation*}
which means that,
\begin{equation}
\label{curvature}
\left| 
\frac{\mathrm{d} \textbf{T}}{\mathrm{d} t} 
\right|^2 = 
\left(
\frac{1}{| \textbf{v} |}
\right)^2
\left(
\left| 
\frac{\mathrm{d} \textbf{v}}{\mathrm{d} t} 
\right|^2
-
\left( 
\frac{\mathrm{d} | \textbf{v} |}{\mathrm{d} t} 
\right)^2
\right)
\end{equation}
%
%
proving we do not need to differentiate 
\textbf{T} to obtain 
$| {\mathrm{d} \textbf{T}} / {\mathrm{d} t} |$.

Now that we know \textbf{T} is 
unit tangent 
to the path traced by a body of velocity 
\textbf{v}, it is clear that 
the derivative of \textbf{T} 
can be 
interpreted 
geometrically.  Namely, 
${\mathrm{d} \textbf{T}} / {\mathrm{d} t}$
is 
the rate 
at which 
the 
tangent turns as that body traces its path. 
We can thus infer that 
$|  
{\mathrm{d} \textbf{T}} / {\mathrm{d} t}
|$ 
must be related 
to that path's {\em curvature}: 
how rapidly 
\textbf{T} turns 
per unit length of 
path. 
Indeed, 
curvature 
$\scalebox{1.3}{$\kappa$} = 
|\, {\mathrm{d} \textbf{T}} / {\mathrm{d} t} \,| 
/ 
| \textbf{v} |$ by definition~\cite{Thomas}.
Combining this formula with~(\ref{curvature}) 
reveals how we can derive 
a geometric 
attribute 
of a body's path 
(curvature $\scalebox{1.3}{$\kappa$}$)
entirely from a physical 
attribute 
of that body (velocity \textbf{v}).
In this way,~(\ref{curvature}) 
confirms  
that a body with velocity \textbf{D}
traces a 
(circular)
path of constant curvature $1 / r$.
By
contrast,
%
a body 
of 
velocity $\textbf{R}$ traces a 
(trochoidal)
path
of
variable
curvature
defined by,
\begin{equation}
\label{kappa}
{\scalebox{1.3}{$\kappa$}}^2 =
\frac{\omega^4}{\left| \textbf{R} \right|^4}
\left(
r^2
-
\frac{r^2 \omega^2 \cos^2 \theta}{G^2 
\left| \textbf{R} \right|^2}
\right)
\end{equation}
wherever $| \textbf{R} | \ne 0$.
Recall that 
$\cos \theta = 0$
at both the highest and lowest points in 
that path.
Hence,~(\ref{kappa}) says 
that 
$\scalebox{1.3}{$\kappa$} = 
r \omega^2 / | \textbf{R} |^2$
%
%
%
%
%
%
%
at such points 
when $| \textbf{R} | \ne 0$.
Given $| \textbf{R} |$ is negatively correlated 
to 
the altitude 
(see~(\ref{relspeed})), 
$\scalebox{1.3}{$\kappa$}$ 
proves 
trochoidal 
waves 
are 
sharper at their peaks than at their  
troughs 
when $| \textbf{R} | \ne 0$: 
a feature mirrored by ocean waves.

\section{Concavity}
\noindent
In the previous section, 
we 
saw how 
changes to 
\textbf{R} 
- 
the
velocity 
of 
a
particle 
tracing 
Gerstner's trochoidal 
wave profile 
- 
generates  
that 
profile's
curvature.
Here, we 
interrogate  
that 
curvature 
to 
expose 
the 
three distinctive  
types of 
trochoidal wave.






Primitive animations 
might 
depict 
the ocean surface 
as a 
sine wave.  
A
sinusoid 
bends 
as 
sharply at 
peaks as it does at 
troughs.
Let us contrast this with a 
trochoidal counterpart:
the curvature 
is 
$r / | ( 1 / G ) - r |^2$ 
at peaks and 
$r / | ( 1 / G ) + r |^2$
at troughs 
(see~(\ref{relspeed}) and~(\ref{kappa})).
Visually then, 
a trochoidal wave 
and 
a sine wave 
actually 
become 
harder to distinguish 
as $r$ becomes 
a smaller fraction of 
$1 / G$.

This is 
revealing.
If a trochoidal 
wave may or may not 
look like a sine wave,  
it follows that 
two Gerstner waves may 
differ greatly 
in shape.
Equation~(\ref{kappa}) suggests 
where we might look for fundamental 
differences between Gerstner waves. 
Consider for instance 
whether 
or not 
the values of 
$r$ and $G$ dictate 
that 
$| \textbf{R} | = 0$. 

Without wishing to pre-empt ourselves,  
these zeros mark points of 
distinguishing geometry.
Let us explain why.
A body 
has 
speed $| \textbf{v} | = 0$ 
if and only if 
that same body has velocity 
$\textbf{v} = \textbf{0}$.  
Where both $\textbf{v} = \textbf{0}$ 
and $| \textbf{v} | = 0$, 
we can define no unit tangent 
$\textbf{T} = \textbf{v} / | \textbf{v} |$ to 
a
body's path.  
Where \textbf{T} 
does not exist,
no 
derivative of 
\textbf{T} exists. 
In short, 
bodies of 
velocity 
$\textbf{v} = \textbf{0}$ 
lie 
at 
points 
where 
$\scalebox{1.3}{$\kappa$} = 
|\, {\mathrm{d} \textbf{T}} / {\mathrm{d} t} \,| 
/ 
| \textbf{v} |$ 
is an indeterminate form. 
%
%
%
%
%
%
%
%
%
%
%
Using~(\ref{relvel}), it is clear that 
such points are found on  
trochoidal waves 
where, 
\begin{equation}
\label{zerovel}
\textbf{R} = \textbf{0} \Longleftrightarrow
( \cos \theta , \sin \theta ) = 
\left( 0 , \frac{1}{r G} \right).
\end{equation}
Figure~\ref{thetadiag} 
illustrates that 
$\cos \theta = 0$
at peaks (where $\sin \theta = 1$) and troughs 
(where $\sin \theta = - 1$)
only. 
Given $G > 0$, 
$\sin \theta = - 1$ is incompatible 
with~(\ref{zerovel}).
Therefore, 
equation~(\ref{zerovel}) 
predicts 
the 
velocity of a 
our 
fluid particle 
vanishes 
if and only if 
$\sin \theta = 1 = 1 / r G$.
Or if you prefer, 
our 
particle's 
velocity 
vanishes 
exclusively at 
the peaks of $r G = 1$ trochoids.


When 
$r G = 1$,
it is 
telling that
fluid particles 
should trace  
trochoidal paths
with 
velocity,
\begin{equation}
\label{cycloid}
\textbf{R} = 
r \omega \left( 1 - \sin \theta \right) \, \textbf{i}
+ r \omega \left( \cos \theta \right) \, \textbf{k} 
\end{equation}
according to~(\ref{relvel}).
Thus, 
Gerstner's 
particles 
bear the 
character of 
particles  
tracing an (upside-down) {\em cycloid}~\cite{Nelson}.  


Equation~(\ref{cycloid}) not only identifies this 
form of trochoid 
as 
a cycloid, 
but can also be used to classify 
the 
$r G = 1$ peaks geometrically. 
Velocity \textbf{R} does this by 
giving us the cycloidal slope $S$ as a ratio of its 
components.  
Although 
we cannot determine 
$S = \cos \theta / ( 1 - \sin \theta )$ 
at 
cycloidal 
peaks 
(where $\theta = a = 2 \pi q + \pi / 2$ 
for $q \in \mathbb{Z}$),  
we can 
evaluate 
the limit of $S$ as $\theta$ approaches $a$.
While this is not 
possible by directly substituting $\theta = a$ 
into $S$, we can apply L'H\^{o}pital's rule to find 
that, 
\begin{equation*}
\lim_{\theta \to a} 
\, S = 
\lim_{\theta \to a} 
\frac{\frac{\mathrm{d}}{\mathrm{d} \theta} ( \cos \theta )}
{\frac{\mathrm{d}}{\mathrm{d} \theta} ( 1 - \sin \theta )}
%
%
=
\lim_{\theta \to a} 
\, \tan \theta 
\end{equation*}
which means that,
\begin{equation}
\label{limits}
\lim_{\theta \to a} 
\, S 
= 
\left\{ 
\begin{array}{ll}
+ \infty, ~~~\mbox{as $\theta \to a^{-}$} \\
- \infty, ~~~\mbox{as $\theta \to a^{+}$}. 
\end{array}
              \right.
              %
              %
\end{equation}
As~(\ref{limits}) 
says, 
the slope of a
cycloidal wave 
is 
positive on one side of 
any  
peak, 
but negative on the other~\cite{Burgiel}.
%
%
When tracing 
a 
cycloid 
then,
a 
fluid particle
makes sharp about-turns at 
peaks, proving 
such peaks 
to be 
{\em vertical cusps}~\cite{Naik}.  
The 
rapid change in direction corresponds 
to an infinitely fast rotation of the tangent, 
which is 
precisely what infinite curvature measures.
%
%


Let us now shift our focus 
to points of zero curvature.  
Given 
the curvature of a trochoidal wave is
$\scalebox{1.3}{$\kappa$} = 
|\, {\mathrm{d} \textbf{T}} / 
{\mathrm{d} t} \,| 
/ 
| \textbf{R} |$ 
by definition, 
it is clear that $\scalebox{1.3}{$\kappa$} = 0$
if and only if both 
$|\, {\mathrm{d} \textbf{T}} / 
{\mathrm{d} t} \,| = 0$ and $| \textbf{R} | \ne 0$.
Geometrically then, 
points of zero curvature 
are those where 
tangent 
$\textbf{T} = \textbf{R} / | \textbf{R} |$
exists 
(since $| \textbf{R} | \ne 0$),  
but 
has stopped 
turning (since $|\, {\mathrm{d} \textbf{T}} / 
{\mathrm{d} t} \,| = 0$).

Using the right hand side of~(\ref{curvature}), 
%
points 
on our wave have
zero curvature 
if and only if 
they satisfy,
\begin{equation}
\label{inflection}
r^2
=
\frac{r^2 \omega^2 \cos^2 \theta}
{G^2 \left| \textbf{R} \right|^2} 
\Longrightarrow 
%
%
%
%
\left( r G - 
\sin \theta \right)^2 = 0
\end{equation} 
without violating our 
$| \textbf{R} | \ne 0$ condition.  


No $r G > 1$ satisfies~(\ref{inflection}), 
as $\sin \theta \le 1$.  
Nor 
can $r G = 1$ satisfy~(\ref{inflection}), 
as~(\ref{zerovel}) confirms that 
$\sin \theta = r G = 1$ 
violates 
our 
$| \textbf{R} | \ne 0$ condition on~(\ref{inflection}).  
Simply put, 
trochoidal 
profiles 
possess 
points of zero 
curvature 
if any only if 
$r G < 1$.  






Equations~(\ref{zerovel}) 
and~(\ref{inflection}) 
reveal the altitude 
of points of notable curvature.  
As 
Figure~\ref{thetadiag} 
illustrates, 
the value of 
$r \sin 
\theta$ defines 
the altitude of 
our 
fluid particle, 
$D$.
Hence, 
equation~(\ref{zerovel}) proves 
that cycloidal 
points of 
infinite 
curvature have altitude 
$z = 1 / G$
while~(\ref{inflection}) proves 
points of zero curvature have 
altitude $z = r^2 G$.

To classify 
points 
of zero curvature 
geometrically, 
we 
use~(\ref{relvel}) to
inspect the trochoidal slope 
$S_T$. 
More specifically, we  
inspect,
\begin{equation}
\label{changeinslope}
{\mathrm{d}} {S_T} / {\mathrm{d} t} = 
\frac{
r^2 \omega - \frac{r \omega}{G} 
\sin \theta 
}{\left( \frac{1}{G} 
- r \sin \theta \right)^2}
= 
\frac{
\omega
\left( 
r^2 - \frac{1}{G} z 
\right)
}{\left( \frac{1}{G} 
- z \right)^2}
\end{equation}
to determine how 
tangent \textbf{T} 
behaves 
as a 
fluid particle 
rises or falls 
through
altitude 
$z = r^2 G$.
Since points of zero curvature lie on the 
$r G < 1$ wave, 
they 
lie 
below trochoidal peaks (where $z = r$) 
and 
above the rest position (where $z = 0$). 

As~(\ref{changeinslope}) shows then, 
${\mathrm{d}} {S_T} / {\mathrm{d} t}$ 
is positive 
at the rest position, 
zero 
at points of zero curvature, and 
negative 
at the $r G < 1$ peaks.  Put simply, 
this tells us 
tangent 
\textbf{T} is 
turning counterclockwise 
as a 
particle 
rises from the rest position 
until it reaches 
an altitude 
of $z = r^2 G$.  At this altitude, the turning 
pauses momentarily.
As a 
particle 
rises 
beyond an altitude of $z = r^2 G$, the turning 
of \textbf{T} resumes, but now in the clockwise 
direction instead.
Tangent \textbf{T} changes its direction of 
rotation at $z = r^2 G$.  Our points of 
zero curvature are thus {\em{inflections}}.

\begin{figure}[h]
\begin{center}
\includegraphics[width=0.42\textwidth]{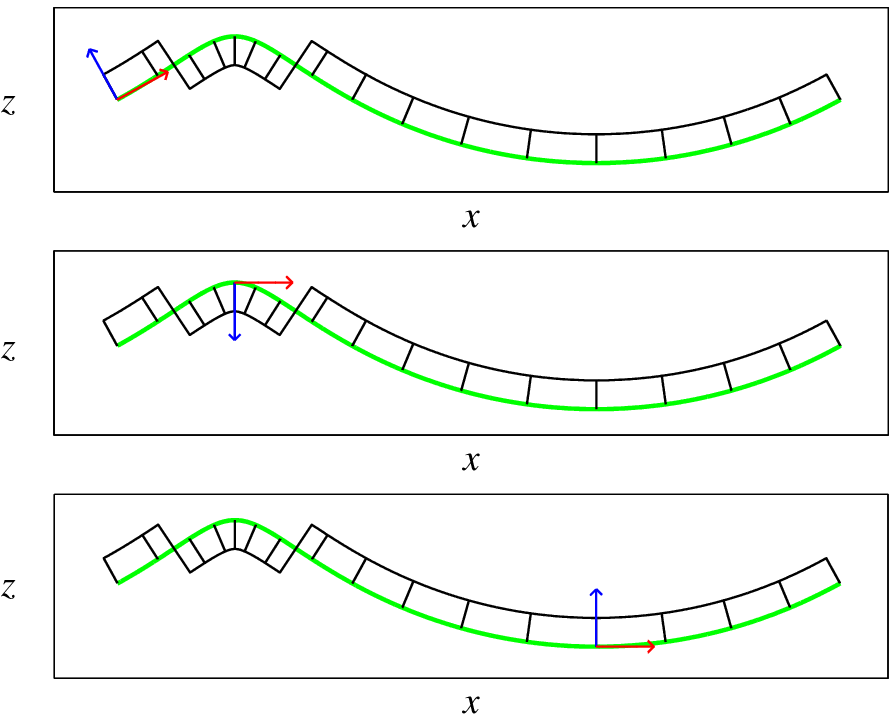}
\hspace{1.75em}
\includegraphics[width=0.245\textwidth]{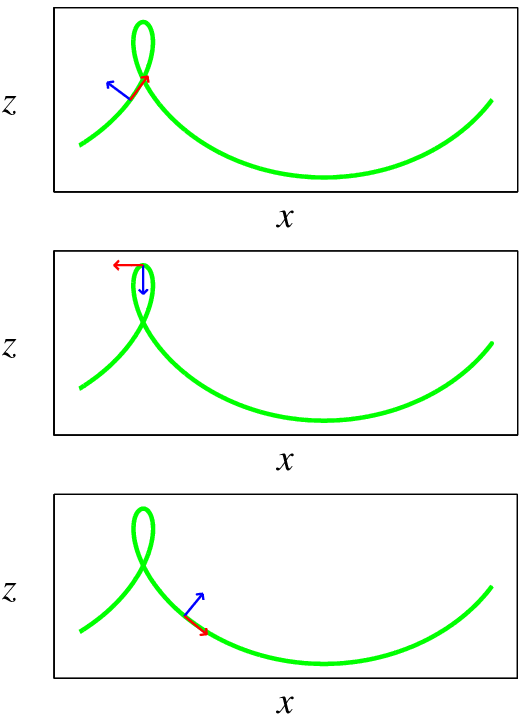}
\caption{Tangent 
$\textbf{T} = \textbf{R} / | \textbf{R} |$ 
(red) and 
corresponding 
normal 
$\textbf{N} = 
( {\mathrm{d} \textbf{T}} / 
{\mathrm{d} t} ) 
/ 
\left| {\mathrm{d} \textbf{T}} / 
{\mathrm{d} t} \right|$ 
(blue) to path (green) form an 
intrinsic coordinate system.  
The associated 
grid (black) is illustrated on the normal's 
side of the path. 
As a 
particle 
advances along 
the 
path, 
tangent \textbf{T} turns toward the side of this path that 
\textbf{N} points to.  
In upper images 
($r G < 1$), \textbf{T} 
is turning 
counterclockwise, 
then 
clockwise, 
then 
counterclockwise again.
In lower images 
($r G > 1$), 
\textbf{T} is turning counterclockwise 
in all images.}
\label{intrinsic}
\end{center}
\end{figure}

In conclusion, 
our 
trochoidal profiles  
do not 
exhibit 
points of infinite curvature unless $r G = 1$, 
nor do they 
exhibit  
points of zero curvature unless $r G < 1$. 
Hence, 
tangent 
$\textbf{T}$ 
turns continuously 
in a single direction 
as a 
fluid particle 
traces profiles where $r G > 1$ 
(see Figure~\ref{intrinsic}).
Given this rotation 
results in 
opposite horizontal tangents 
at 
peaks and troughs 
when $r G > 1$
(see~(\ref{relvel})), 
we conclude the 
profile 
{\em{self-intersects}}. 

\section{Crossings}
In the previous section, we saw how  
the 
time-evolution of 
$\textbf{T} = 
\textbf{R} / | \textbf{R} |$ 
- the
tangent to a trochoidal wave - 
can forecast that wave's {\em concavity}.
As such, 
Gerstner's model predicts the profile of a 
breaking (unstable) wave  
to cross itself.  
Such profiles are physically 
unrealistic~\cite{Okamoto}.  
Where 
self-intersections can be 
precisely located though, 
users 
can truncate 
a 
wave 
profile 
accordingly.  
 

When $r G > 1$, our trochoidal wave profile 
exhibits a loop and a point of self-intersection: 
a {\em node}.  
%
%
%
When 
such profiles 
are rendered 
using~(\ref{xandy}), 
a 
node becomes a point in space 
which 
could be 
occupied 
by 
two 
fluid particles 
at once.

To 
avoid this
interpretation,  
let us instead 
view a 
trochoidal 
profile as 
the path traced by just one 
fluid 
particle, 
$D$.  
Equation~(\ref{relvel}) defines the 
velocity of $D$ as it 
traces 
the 
profile 
to be \textbf{R}.
By 
integrating 
\textbf{R}
with respect to 
$t$, we 
therefore 
recover 
the trochoidal profile that $D$ 
traces over time:
\begin{equation}
\label{relvelint}
\int \textbf{R} \, \mathrm{d} t
= 
\left( 
r \cos \theta + \frac{\omega}{G} t +  
C_1
\right) \textbf{i} 
+ 
\left(
r \sin \theta + 
C_2
\right)
\textbf{k}
\end{equation}
where  
$C_1$ and $C_2$ are constants of integration.  
By setting 
$t$ equal to zero, we obtain 
$C_1 = x_0 - r \cos ( G b )$ and 
$C_2 = z_0 - r \sin ( G b )$, where $x_0$ 
and $z_0$ are the initial coordinates of $D$.


Featured in 
Figure~\ref{intrinsic}, trochoidal wave profiles 
appear to exhibit a vertical line of symmetry 
through any chosen trochoidal peak, say $p$.  
To prove 
this line 
exists,  
we 
suppose $D$ to be at 
peak $p$ 
when $t = 0$.  
Hence, $D$ has 
initial coordinates 
$\left( x_0 , z_0 \right) = 
( b , r )$
and 
an 
initial 
phase of 
$\theta = G b 
= \pi / 2 + N 2 \pi$ for some 
$N \in \mathbb{Z}$.  Using $N = 0$, 
let us compare the positions of $D$ 
at $t = f$ and $t = - f$. 


Given $G b = \pi / 2$, four values 
follow:
$\sin ( G b ) = 1$, $\cos ( G b ) = 0$, 
$C_1 = x_0$ 
and 
$C_2 = 0$. 
By applying  
sum formulae 
to the trigonometric 
functions of~(\ref{relvelint}) and 
substituting these four values into 
the results,  
we find that $D$ has general coordinates, 
\begin{equation} 
\label{beginatpeak}
\chi = - r \sin ( \omega t ) + 
\frac{\omega t}{G}
+ 
b
~~~\mbox{ , }~~~
z = r \cos( \omega t )
\end{equation}
in this example.  
As Galilean relativity dictates, 
the horizontal coordinate of $D$ 
above, 
$\chi$, is equal to $x - \omega t / G$.

Since cosine 
is an even function, 
we can conclude that 
$D$ has the 
same height at $t = \pm f$ 
when $D$ begins at a peak.
In our example, 
$\chi - x_0$ at $t = f$ is 
the negative of $\chi - x_0$ at $t = -f$ 
since sine is an odd function.  
As 
$f$ is arbitrary, 
these properties prove 
that a vertical line of symmetry 
does 
pass through peak $p$.

Any nodes must obey this 
line of symmetry.  
Let us assume that Figure~\ref{intrinsic} 
is not atypical of the $r G > 1$ case: 
there exists 
a single 
node between 
$\chi = x_0 - \lambda / 2$ and $\chi = 
x_0 + \lambda / 2$ when $r G > 1$, 
where $\lambda = 2 \pi / G$ is the trochoidal 
wavelength.  
That 
node 
could not obey our line of symmetry without 
lying upon it.
As such, our node will share the 
$\chi$ value of $p$.



When $D$ begins at peak $p$ then, 
$D$ will 
reach different 
points 
on the wave 
when $t = f$ and $t = -f$.  
Except when $D$ reaches our node.  
Hence, our node is the only point on 
the wave where $\chi$ at $t = f$ 
equals
$\chi$ at $t = - f$ for $f \ne 0$, 
meaning, 
%
%
%
%
%
%
%
%
%
%
%
\begin{equation}
\label{nodecond}
\frac{\omega}{G} f 
= 
r \sin ( \omega f ) 
\Longleftrightarrow
\frac{\sin ( \omega f )}{\omega f}
= \frac{1}{r G}
\end{equation}
holds for no point other than our node if 
$f \ne 0$, according to~(\ref{beginatpeak}).
\begin{remark}
Readers may be interested to know that 
$\sin ( \omega f ) / 
\omega f$ is 
the 
{\em cardinal sine} 
({\em sinc}) 
function of $\omega f$.  
The function is undefined for $\omega f = 0$.   
Consequently, we need to define the function's 
value at $\omega f = 0$ based on the limit as 
$\omega f$ approaches 0.  This limit is 1.
%
%
By the nature of sinc functions, it follows 
that 
$\sin ( \omega f ) / 
\omega f < 1$ for all non-zero values of 
$\omega f$.  Therefore, 
as we would expect, 
we cannot solve~(\ref{nodecond}) 
for any $r G \le 1$.
\end{remark}
In general, 
solving~(\ref{nodecond}) for $f$ is not 
straightforward.  One 
specific 
solution 
is self-evident though.  By comparing the 
numerators and denominators of~(\ref{nodecond}), 
we 
can see 
that 
$\sin ( \omega f ) = 1$ 
(and therefore $\cos ( \omega f ) = 0$)
when $\omega f = r G$. 
Substituting these 
expressions 
into~(\ref{beginatpeak}),
we 
can 
thus
establish 
that 
the node which lies on the line of 
symmetry through $p$ 
has 
coordinates $( \chi , z ) = ( x_0 , 0 )$ 
when $\omega f = r G$.

We can generalise this approach by 
adapting equation~(\ref{nodecond}) 
into the form,
\begin{equation}
\label{nodegeneral}
\frac{\sin ( \omega f )}{\omega f}
= \frac{1}{r G} 
\Longleftrightarrow
\frac{\sin ( \omega f )}{\omega f}
= \frac{A}{A r G}
\end{equation}
where 
$A \ne 0$ is a value to be determined.
When we choose $A$ to equal 
$\sin ( \omega f )$,
equation~(\ref{nodegeneral}) 
tells us that 
$\omega f = A r G$.  By substituting these 
expressions into~(\ref{beginatpeak}), 
we find that a node which lies on the line 
of symmetry through $p$ has coordinates,
\begin{equation}
\label{nodecoords}
\left( \chi , z \right) = 
\left( x_0 , r \cos ( A r G ) \right)
\end{equation}
when $\omega f = A r G$.

To be clear,~(\ref{nodecoords}) 
defines our node's location 
for a chosen value of $A r G$.  
By equating the 
denominators of~(\ref{nodegeneral}), 
we thus obtain a 
value for $\omega f$.
By equating the 
numerators of~(\ref{nodegeneral}), 
we obtain a value for $A \in [ -1 , 1 ]$.
Then we obtain a value for $r G > 1$
by dividing our initially chosen 
$A r G$ value by $A$.


Unlike inflections and cusps, 
nodes can lie 
below the rest 
positions 
of Gerstner waves.
Equation~(\ref{nodecoords}) says that nodes 
have altitude $z < 0$ 
when,
\begin{equation*}
\frac{\pi}{2} < A r G < \pi
\Longleftrightarrow
\frac{\pi}{2 A} < r G < \frac{\pi}{A}
\end{equation*}
%
%
noting 
$\omega f$ (which we set equal to $A r G$), 
the change in $\theta$ over 
$f$ units of time, 
need not 
exceed $\pi$ for our purposes.
\begin{table}[ht]
\caption{Examples of solutions 
to~(\ref{nodegeneral}) and~(\ref{nodecoords}), 
where $( x_0 , z )$ 
mark our 
self-intersection's 
coordinates.  One possible value for 
$x_0$ is $b$, 
a value we can determine using 
$G b = \pi / 2$ 
and 
$r G$ values.} 
\centering 
\begin{tabular}{c c c c c} 
\hline 
$A$ & $\omega f$ & $r G$ & $x_0$ & $z$ \\ [0.5ex] 
\hline  
$\frac{1}{2}$ & $\frac{5 \pi}{6}$ & $\frac{5 \pi}{3}$ & $\frac{3 r}{10}$ & $- \frac{\sqrt{3} r}{2}$ \\
[0.75ex]
1 & $\frac{\pi}{2}$ & $\frac{\pi}{2}$ & $r$ & 0 \\ 
$\frac{\sqrt{3}}{2}$ & $\frac{\pi}{3}$ & $\frac{2 \pi}{3 \sqrt{3}}$ & $\frac{3 \sqrt{3} r}{4}$ & $\frac{r}{2}$ \\ [0.75ex]
$\frac{\sqrt{2}}{2}$ & $\frac{\pi}{4}$ & $\frac{\pi}{2 \sqrt{2}}$ & $\sqrt{2} r$ & $\frac{\sqrt{2} r}{2}$ \\ 
[1ex]
\hline 
\end{tabular}
\label{table:node} 
\end{table}

\section{Cycloids}
In the previous section, we 
saw 
how 
the 
profiles of 
Gerstner's 
water 
waves 
exhibit (unrealistic) loops once 
they become overly steep. 
Under Gerstner's model, unstable  
waves manifest as {\em{prolate}} (extended) 
trochoids, 
while stable waves manifest 
as cycloids 
or 
{\em{curtate}} (flattened) trochoids. 

By convention, trochoidal curves are 
generated mechanically.
As a circular wheel 
undergoes 
{\em{pure rolling}} 
(spinning as much as it moves forwards)
along 
a horizontal plane, 
a point 
attached to the 
wheel traces a trochoidal curve.  When 
the point is 
fixed 
to the 
wheel's 
rim, 
the resulting curve is a cycloid.  
When the point is closer to / farther from 
the wheel's 
centre, the result is a shorter / longer 
curve: 
a curtate / prolate trochoid.  
Determining 
just 
how much shorter or longer is the subject of 
this section. 


As we shall go on to explain, some  
trochoidal curves are easier to measure 
than others.  In understanding why this 
might be, 
let us briefly discuss 
an unconventional 
interpretation 
of trochoidal curves. 
Under Phillips' setup, 
all 
three forms of 
wave 
are 
generated by 
fluid particles 
that behave as 
though 
they were points 
on the rims of circular 
wheels of radii $r$. 
As the wheels roll along a horizontal plane, 
it is 
the mode of rolling 
which determines the form of wave 
profile the 
particle 
traces. 
%


There are three modes of rolling.
First, 
suppose 
$| \textbf{D} |$, the (tangential) 
speed of the 
fluid particle, 
matches $| \textbf{P} |$, the speed 
of the wave.  Then the arc 
length travelled by a point on our wheel 
(relative to the axle) 
matches the distance travelled by the axle.
Our wheel spins as much as it moves forwards: 
it 
exhibits pure rolling.  
It turns out that 
$| \textbf{D} | = | \textbf{P} |$ 
corresponds 
precisely 
to the case where 
$r G = 1$, i.e. 
pure rolling generates the cycloidal wave.



Secondly, 
suppose 
$| \textbf{D} |$, the (tangential) 
speed of the 
fluid particle, 
exceeds $| \textbf{P} |$, the speed 
of the wave.  Then the arc 
length travelled by a point on our wheel 
(relative to the axle) 
exceeds the distance travelled by the axle.
Our wheel spins more than it moves forwards: 
it is said to be {\em{slipping}}.
It turns out that 
$| \textbf{D} | > | \textbf{P} |$ 
corresponds 
precisely 
to the case where 
$r G > 1$, i.e. 
slipping generates the prolate 
trochoidal wave.



Third, 
suppose 
$| \textbf{D} |$, the (tangential) 
speed of the 
fluid particle,
is less than $| \textbf{P} |$, the speed 
of the wave.  Then the arc 
length travelled by a point on our wheel 
(relative to the axle) 
is less than the distance travelled by the axle.
Our wheel spins less than it moves forwards: 
it is said to be {\em{skidding}}.
It turns out that 
$| \textbf{D} | < | \textbf{P} |$ 
corresponds 
precisely 
to the case where 
$r G < 1$, i.e. 
skidding generates the curtate 
trochoidal wave.


%
%

Whichever you prefer, 
both interpretations prove that the three 
forms of trochoidal wave profile are generated 
in three materially different ways.  
Perhaps unsurprisingly then, 
no 
single
closed-form expression 
defines the arc length of all three 
trochoidal forms.

\begin{figure}[h]
\begin{center}
\includegraphics[width=0.4\textwidth]{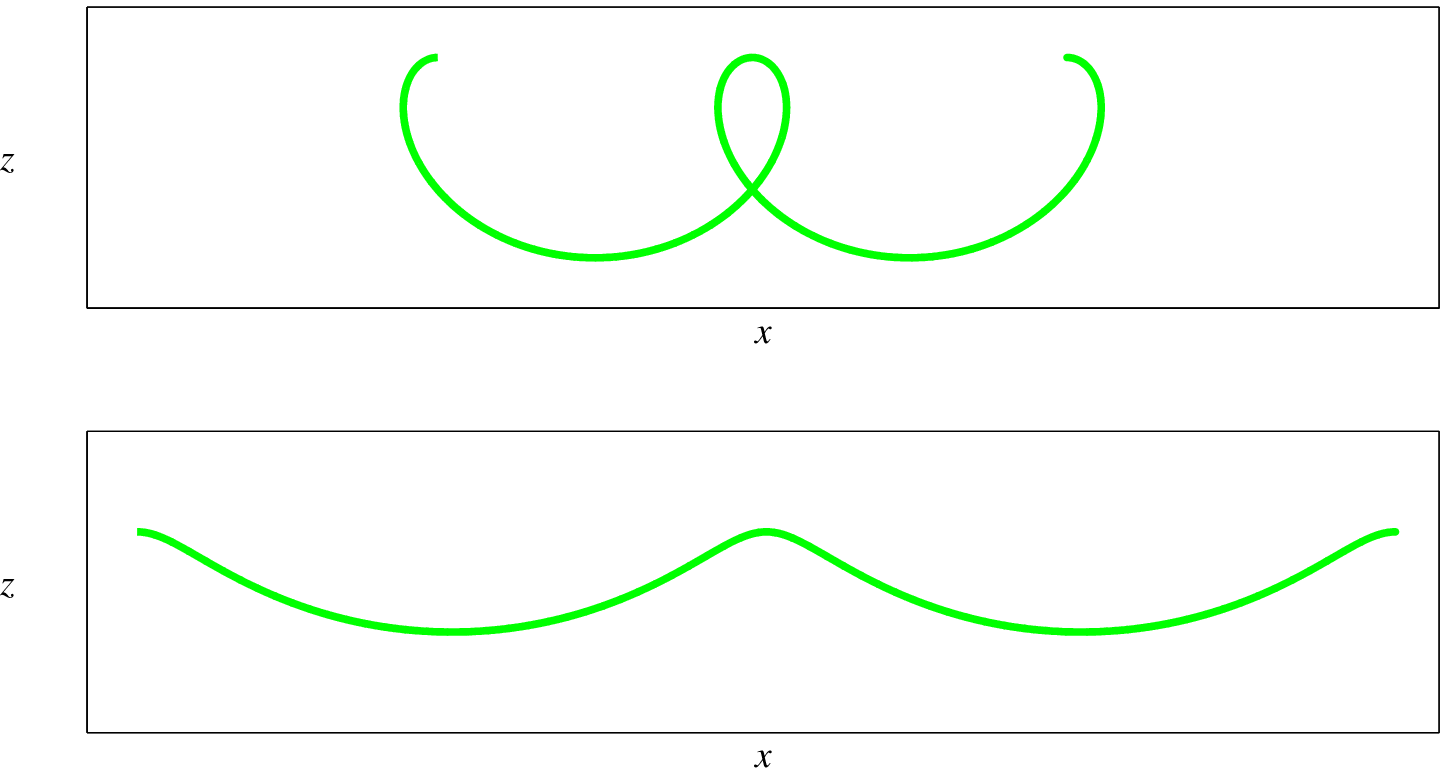}
\caption{Two curves of equal arc length: 
a prolate trochoidal curve (top image: 
$r = 1 / 2$ and $G = 4$), and a curtate 
trochoidal curve ($r = 1 / 4$ and $G = 2$). 
A 
particle 
traces either in  
$4 \pi / \omega$ units of time.}
\label{arcprolcurt}
\end{center}
\end{figure}

Since the arc length of a trochoidal wavelength 
is equal regardless of which points we choose 
as endpoints, let us measure the arc length 
peak-to-peak.  
Using~(\ref{beginatpeak}) 
in the standard formula yields: 
\begin{equation*}
L = 
\int
^{\frac{2 \pi}{\omega}}_0 
\sqrt{ 
\left( \frac{\mathrm{d} \chi}{\mathrm{d} t} \right)^2
+ 
\left( \frac{\mathrm{d} z}{\mathrm{d} t} \right)^2 
}
\, \mathrm{d} t
= \int^{\frac{2 \pi}{\omega}}_0 
| \textbf{R} | \, \mathrm{d} t
\end{equation*}
i.e. 
the 
arc length of a path 
is equal to 
the 
speed of a point tracing that path, 
integrated over 
time.
This 
mirrors the classical 
relation: Distance $=$ Speed $\times$ Time.
Using~(\ref{relspeed}), we obtain:
\begin{equation}
\label{arclength}
L = 
\omega 
\int
^{\frac{2 \pi}{\omega}}_0 
\sqrt{
r^2 + \frac{1}{G^2} - \frac{2 r}{G} \cos ( \omega t )
}
\, \mathrm{d} t.
\end{equation}
Suppose the value of $r G$, which determines 
the form of our trochoidal wave, is equal to $h$.  
If we substitute $r / h$ for $1 / G$ 
into~(\ref{arclength}), we 
can extract a common  
factor of $r$:  
\begin{equation*}
L
= 
r 
\omega 
\int
^{\frac{2 \pi}{\omega}}_0 
\sqrt{
1 + \frac{1}{h^2} - \frac{2}{h} \cos ( \omega t )
}
\, \mathrm{d} t. 
\end{equation*}
Next, we can 
exploit a half-angle formula,
$1 - \cos ( 2 \psi ) = 2 \sin^2 \psi$ to 
determine that, 
\begin{equation}
\label{arclen3}
L
= 
2 r 
\int
^{T = {\pi}}_{T = 0} 
\sqrt{
1 + \frac{1}{h^2} - \frac{2}{h} + 
\frac{4}{h} \sin^2 
T
}
\, \mathrm{d} T
\end{equation}
where we made the substitution 
$T = \omega t / 2$. 
In the cycloidal case,
it 
is clear that 
we can integrate our expression directly 
since~(\ref{arclen3}) becomes,
\begin{equation}
\label{arclencyc}
L
= 
4 r 
\int
^{T = \pi}_{T = 0} 
| \sin T | 
\, \mathrm{d} T
= 
4 r 
\int
^{\pi}_{0}
\sin T 
\, \mathrm{d} T
= 
8 r
\end{equation}
when $h = 1$. 
As~(\ref{arclencyc}) 
says, 
the length 
of a cycloidal wave 
cycle 
depends only its 
height.
The length of any other trochoidal wave proves 
harder to compute, 
the 
integral that 
we must solve in~(\ref{arclen3}) 
being 
{\em{elliptic}}, as we can show more clearly 
by expressing it in the following form:
\begin{equation}
\label{arclenellip}
L
= 
2 r 
\sqrt{Q} 
\int
^{T = {\pi}}_{T = 0} 
\sqrt{
1 - m
\sin^2 
T
}
\, \mathrm{d} T
\end{equation}
where $Q = 1 - ( 2 / h ) + 1 / ( h^2 )$ and 
$m = - ( 4 / h ) / Q$, noting $m$ assumes 
$Q \ne 0$ (equivalent to $h \ne 1$).  
It is the classical elliptic 
form of~(\ref{arclenellip}) which defines the 
arc length of all non-cycloidal trochoids.
\begin{remark}
Equation~(\ref{arclenellip}) is 
a complete 
elliptic integral of the second kind  
that we have reduced to one of 
Legendre's three {\em{canonical}} (normal) 
forms. The integral can be evaluated numerically 
using 
the Arithmetic-Geometric Mean (AGM) technique.
Approximate analytical solutions have been 
proposed~\cite{Chachiyo}.
\end{remark}
While~(\ref{arclenellip}) cannot be 
solved analytically, 
we can still make certain observations 
about our non-cycloidal waves.  
Consider two trochoidal waves: one prolate 
(where $r = r_1$ and $G = G_1$ such that 
$r_1 G_1 > 1$) and one curtate 
(where $r = r_2$ and $G = G_2$ such that 
$r_2 G_2 < 1$).  Without needing to 
solve~(\ref{arclength}), it is clear that 
our curtate and prolate waves have equal 
peak-to-peak arc lengths when:
\begin{equation*}
\frac{2 r_1}{G_1} = \frac{2 r_2}{G_2} ~~\mbox{ and }~~ 
r_1^2 + \frac{1}{G_1^2} = r_2^2 + \frac{1}{G_2^2}
\end{equation*}
which we solve simultaneously to obtain
the condition: 
$r_1 / G_1 
= 
r_2 / G_2$. 
Despite their distinctive geometry, 
the peak-to-peak profiles of 
an 
extended wave 
and flattened wave 
are 
equally long
(e.g. see Figure~\ref{arcprolcurt}) when,
\begin{equation*}
\frac{r_1 \lambda_1}{2 \pi} = 
\frac{r_2 \lambda_2}{2 \pi}
\Longleftrightarrow
2 \pi r_1 \lambda_1 =
2 \pi r_2 \lambda_2.
\end{equation*}
A point 
on 
the 
generating 
wheel's rim travels 
a distance of $2 \pi r$ 
with respect to the axle, 
for every $\lambda$ the 
axle travels horizontally. 
In short, 
the product of a 
generating 
wheel's 
rotation and translation 
is equal for both 
trochoids.

\section{Overview: Richness and realism}
Over two hundred years old, 
Gerstner's 
equations 
model 
water
waves 
as trochoidal. 
Any such form 
is best 
understood 
by 
analysing 
\textbf{R}, 
the velocity of a 
particle
as it  
traverses the wave.  How this 
velocity 
changes over time 
relates 
the wave structure.

By convention, velocity 
is 
considered 
a physical vector quantity.  
Nonetheless,
a body's 
velocity 
\textbf{v}
can be expressed  
as a product of 
its 
speed 
$| \textbf{v} |$
and 
geometrical 
vector \textbf{T}:
the tangent to that body's path. 
To judge how velocity changes over 
time then, we apply 
the product rule of 
differentiation 
to 
${\textbf{v}} = | \textbf{v} | \textbf{T}$:  
\begin{equation}
\label{accelannotate}
\underbrace{
\frac{\mathrm{d} \textbf{v}}{\mathrm{d} t} 
}_{\substack{
\vphantom{Acceleration Ctraipht}
\text{Total Acceleration}}}
=
\underbrace{
\frac{\mathrm{d} 
| \textbf{v} |}{\mathrm{d} t} 
\textbf{T} 
}_{\substack{
\vphantom{Acceleration Ctraipht}
\text{Tangential Acceleration}}}
+ 
\underbrace{
| \textbf{v} |
\frac{\mathrm{d} \textbf{T}}{\mathrm{d} t}.
}_{\substack{
\vphantom{Acceleration Ctraipht}
\text{Normal Acceleration}}} 
\end{equation}
As 
illustrated in 
Figure~\ref{intrinsic},~(\ref{accelannotate}) 
decomposes 
acceleration 
into 
perpendicular components:
one 
in the 
direction of motion, 
one 
toward the centre of curvature~\cite{UniTexas,Widnall}.

Each component 
plays 
a distinctive 
role  
in our analysis.
%
Interrogating 
${\mathrm{d} | \textbf{v}} |  
/ {\mathrm{d} t}$ 
can 
reveal not just 
how much a body of velocity \textbf{v}
is 
quickening 
or slowing but whether two bodies 
of 
velocity \textbf{v} 
are drawing closer 
or pulling apart.
This can influence the 
distribution 
of bodies 
traversing 
a single 
path.
When $\textbf{v} = \textbf{R}$,
fluid particles 
less than 
half a 
cycle 
apart 
are 
thus 
closer when 
one is at a peak than when one is at a trough 
(see Figure~\ref{frame}).

Interrogating the 
${\mathrm{d} \textbf{T}} 
/ {\mathrm{d} t}$ of~(\ref{accelannotate}) 
tells us how 
much 
a 
path of 
tangent \textbf{T}
is turning 
at time $t$. 
How \textbf{T} evolves 
with $t$ 
thus 
defines 
the 
geometry
of a body's path.  
When 
$\textbf{T} = \textbf{R} / | \textbf{R} |$, 
this 
path is 
trochoidal. 
As~(\ref{kappa}) shows, 
$\textbf{T} = \textbf{R} / | \textbf{R} |$
turns more rapidly at 
peaks than it does at trochoidal troughs, 
proving the peaks to be sharper. 

Despite 
sharing 
these 
properties, 
any two trochoidal waves may have geometric 
features which distinguish them from one 
another.  
Depending on 
a user's choice of 
parameter values,
rendered 
waves 
exhibit one of 
three 
features: 
nodes
(when $r G > 1$), cusps 
(when $r G = 1$), or  
inflections (when $r G < 1$).



\begin{figure}[h]
\begin{center}
\includegraphics[width=0.48\textwidth]{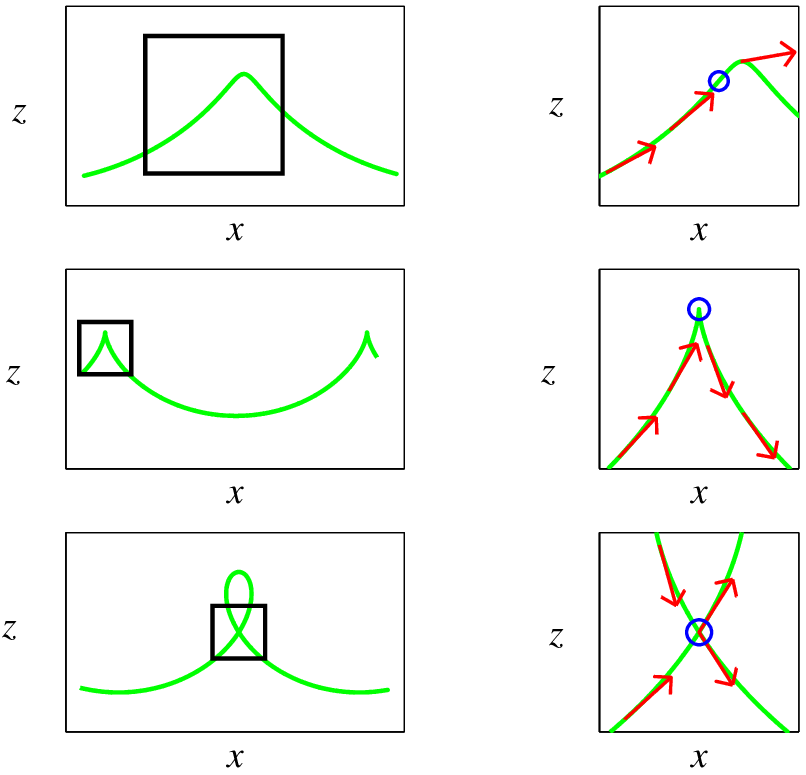}
\caption{Gerstner waves when $r G < 1$ (top), 
$r G = 1$ (middle), and $r G > 1$ (bottom).
Zooming out (left) and in (right) on 
circled points: 
inflection 
(top), 
cusp 
(middle), 
and 
node 
(bottom). 
Tangent 
(unit velocity) vectors (red) shown
local to each point.}
\label{zoomin}
\end{center}
\end{figure}

Let us 
characterise 
each 
of these 
features 
by 
the behaviour 
of tangents. 
Wherever two 
branches 
of a trochoidal 
profile 
cross one another, 
we can define 
two
different
 tangents: 
one for each branch (as in Figure~\ref{zoomin}).  
Thus, a single point on a trochoidal wave 
cycle 
- 
a node - 
exhibits 
multiple 
distinct 
tangents when $r G > 1$.

If users instead lower the value of 
$r G$ to one, the  profile becomes a cycloid:  
nodes 
degenerate into 
cusps, as 
a node's 
non-coincident tangents  
come to coincide at 
a cycloidal peak.
%
%
%
Velocity $\textbf{R}$
vanishes 
at 
these cusps such that they 
exhibit no 
unit
tangent 
$\textbf{T} = \textbf{R} / | \textbf{R} |$ 
(as in Figure~\ref{zoomin}). 




Visually, Gerstner waves 
approximate sinusoids 
when 
$r G \ll 1$.
Like tangents of the latter, 
tangents of the former 
instantaneously
stop turning   
at some height 
when $r G < 1$.
Tangent 
\textbf{T} 
turns 
clockwise above, 
and   
counterclockwise 
below, 
this height 
in Figures~\ref{intrinsic} and~\ref{zoomin}.    
Points 
at this height 
- inflections - 
exhibit one non-rotating tangent. 

\begin{table}[ht]
\caption{Properties of 
Trochoidal waves.}
\centering 
\begin{tabular}{c | c c c} 
\hline 
$ $ & $r G < 1$ & $r G = 1$ & $r G > 1$ \\ [0.5ex] 
\hline  
Arc length & See~(\ref{arclenellip}) & $8 r$ &
See~(\ref{arclenellip}) \\ [0.75ex]
Key point $p$ &
Inflection & Cusp & Node  \\ 
[0.75ex]
Altitude of $p$ &
$r^2 G$ & $r$ & $r \cos ( A r G )$ \\ [0.75ex]
Tangents 
at $p$ &
1 & 0 & 2 \\
[0.75ex]
$| {\mathrm{d} \textbf{T}} / {\mathrm{d} t} |$ at $p$ & 
0
& 
N/A
& 
Non-zero \\
[1ex]
\hline 
\end{tabular}
\label{table:props} 
\end{table}

When users choose values that dictate 
$r G \le 1$, 
a stable wave is modelled.
Otherwise, a 
breaking wave is
represented.
Whether viewed through a physical lens 
or through a geometric lens, let us show that 
Gerstner waves thus obey an intuitive stability 
threshold.
Although 
not obvious at first glance, $r G$ is equal 
to a fraction:
\begin{equation}
\label{stability}
r G =  
\frac{r \left| \omega \right|}
{\left| \omega \right| / G} 
= 
\frac{\left| \textbf{D} \right|}
{\left| \textbf{P} \right|}
\end{equation} 
which exceeds one if and only if 
$| \textbf{D} | > | \textbf{P} |$.
Hence, 
equation~(\ref{stability}) 
says 
Gerstner's model 
fits with real world 
observation: 
gravity waves 
%
%
%
will 
break 
once 
constituent 
particles 
move faster than the waves do.

While Gerstner's profile of a breaking wave 
may not be realistic, we can 
nevertheless 
constructively 
interpret~(\ref{stability})
geometrically.  Gerstner's model predicts 
that a gravity wave breaks once,
\begin{equation}
\label{stabgeom}
r G > 1 
\Longleftrightarrow
r \frac{2 \pi}{\lambda} > 1 
\Longleftrightarrow
2 r > \frac{\lambda}{\pi}.
\end{equation}
Put simply, 
the right-most  
inequality in~(\ref{stabgeom}) 
predicts that 
gravity 
waves 
grow 
unstable 
once 
their height ($2 r$) 
exceeds a fraction of their wavelength.
Thus, Gerstner's model 
imitates 
Nature: 
a
gravity 
wave will break if the surface slopes too 
steeply~\cite{Massel}.

%
%

%
%
%
%
%
%
%
%
%
%
%

\section{Discussion: Galilean transformations}
Gerstner's 
fluid particles 
highlight
a 
key 
distinction between 
velocity and acceleration. 
As 
\textbf{D} and 
$\textbf{R} = \textbf{D} - \textbf{P}$ 
demonstrate, 
the velocity 
of a body is relative: a change 
of reference frame 
can 
change the 
velocity.  By contrast, 
equation~(\ref{Sderiv}) 
reveals 
${\mathrm{d} \textbf{D}} / {\mathrm{d} t} = 
{\mathrm{d} \textbf{R}} / {\mathrm{d} t}$.  
Swapping 
one 
non-accelerating and 
non-rotating  
reference 
frame 
for 
another 
- what is 
called 
a 
{\em Galilean transformation} -
cannot 
change 
the size or direction of any  
acceleration 
vector.  A body's acceleration is 
absolute~\cite{Sokal}.

However, 
even 
acceleration
vectors 
which are equal 
may
decompose
differently 
(e.g. see Figure~\ref{accelcomp}). 
While 
such 
vectors decompose identically 
in 
a 
single 
coordinate system, 
differences may emerge when we 
instead 
use 
the {\em intrinsic 
coordinates} of a body's motion.  
By applying the formula 
%
in~(\ref{accelannotate}) to 
${\mathrm{d} \textbf{D}} / {\mathrm{d} t} =
{\mathrm{d} \textbf{R}} / {\mathrm{d} t}$ 
for instance, 
such coordinates yield, 
\begin{equation}
\label{dDdt&dRdt}
\underbrace{
| \textbf{D} | 
\frac{\mathrm{d} \textbf{T}_D}{\mathrm{d} t} 
}_{\substack{\text{ 
\vphantom{Acceleration straight}
Acceleration of water observed}
\\
\text{
\vphantom{Acceleration straight}
from vessel of zero velocity}
}
}
= 
\underbrace{
\frac{\mathrm{d} | \textbf{R} |}{\mathrm{d} t} 
\textbf{T}_R
+ 
| \textbf{R} |
\frac{\mathrm{d} \textbf{T}_R}{\mathrm{d} t}
}_{\substack{\text{ 
\vphantom{Acceleration straight}
Acceleration of water observed}
\\
\text{
\vphantom{Acceleration straight}
from vessel of velocity \textbf{P}}
}
}
\end{equation}
when 
unit 
tangents 
$\textbf{T}_D = \textbf{D} / | \textbf{D} |$ 
and $\textbf{T}_R = \textbf{R} / | \textbf{R} |$ 
exist, 
noting
${\mathrm{d} | \textbf{D} |} / {\mathrm{d} t} = 0$.  
Despite 
${\mathrm{d} \textbf{D}} / {\mathrm{d} t}$ 
and 
${\mathrm{d} \textbf{R}} / {\mathrm{d} t}$ 
being equal, 
equation~(\ref{dDdt&dRdt}) shows that 
their tangential components 
are 
generally 
unequal, as 
therefore 
are their 
normal 
components.
While 
{\em doppelg\"{a}ngers} of sorts,
${\mathrm{d} \textbf{D}} / {\mathrm{d} t}$ 
and 
${\mathrm{d} \textbf{R}} / {\mathrm{d} t}$ 
are 
thus 
not interchangeable.  
The tangential component of just one 
- that of 
${\mathrm{d} \textbf{R}} / {\mathrm{d} t}$ 
-  
can 
convey 
the 
spatial distribution 
of Gerstner's 
fluid particles  
and
the 
normal 
component of 
just one -
that of 
${\mathrm{d} \textbf{R}} / {\mathrm{d} t}$
-
can 
convey the shape of 
Gerstner waves.

According to~(\ref{dDdt&dRdt}),
a
fluid 
particle 
that 
exhibits constant speed along a trajectory 
of constant curvature
may simultaneously be seen to 
quicken and slow on a trajectory 
of 
sharp and gentle turns.
In this way, 
Gerstner's 
particle 
reflects 
the Galilean principle:
total 
acceleration is absolute, but 
it 
can 
be  
redistributed.
Figure~\ref{accelcomp} illustrates how. 
Galilean transformations can rescale  
and / or redirect 
a body's 
tangential
acceleration and 
normal
acceleration. 

Curiously, 
a 
particularly 
consequential 
Galilean 
transformation
occurs 
at points where 
neither 
a 
particle's  
normal 
nor 
its 
tangential 
acceleration is resized or redirected.  
A case 
we encountered 
in~(\ref{kappa}), 
${\mathrm{d} | \textbf{R}} | / 
{\mathrm{d} t} = 0$ 
at 
trochoidal peaks and troughs 
(when $| \textbf{R} | \ne 0$). 
Consulting~(\ref{dDdt&dRdt}), 
this case 
implies,
\begin{equation}
\label{dDdt&dRdt2}
| \textbf{D} |
\left|
\frac{\mathrm{d} \textbf{T}_D}{\mathrm{d} t}
\right|
= 
| \textbf{R} |
\left|
\frac{\mathrm{d} \textbf{T}_R}{\mathrm{d} t}
\right|
\Longrightarrow
| \textbf{D} |^2 \,
{\scalebox{1.3}{$\kappa$}}_D
= 
| \textbf{R} |^2 \,
{\scalebox{1.3}{$\kappa$}}_R
\end{equation}
at the highest and lowest points in 
a 
fluid particle's  
path, whether it 
is observed 
to 
trace 
(circular)
orbits 
of curvature 
${\scalebox{1.3}{$\kappa$}}_D$ 
or 
(trochoidal)
wave profiles 
of curvature 
${\scalebox{1.3}{$\kappa$}}_R$.

At 
a 
fluid particle's 
highest and lowest points then, 
equation~(\ref{dDdt&dRdt2}) tells us that 
$| \textbf{R} | > | \textbf{D} |$ if and 
only if ${\scalebox{1.3}{$\kappa$}}_R < 
{\scalebox{1.3}{$\kappa$}}_D$ and that 
$| \textbf{R} | < | \textbf{D} |$ if and 
only if ${\scalebox{1.3}{$\kappa$}}_R > 
{\scalebox{1.3}{$\kappa$}}_D$.
To a sailor who matches 
the wave velocity, 
particles 
appear 
faster 
at their lowest points 
than 
they do 
to the sailor at rest 
(see~(\ref{instvel}) 
and~(\ref{relspeed})).
That is, $| \textbf{R} | > 
| \textbf{D} |$ since 
$| \textbf{R} | = 
| ( 1 / G ) + r | 
| \omega |$ 
while 
$| \textbf{D} | = 
r | \omega |$.
As~(\ref{dDdt&dRdt2}) predicts 
then, 
the troughs of 
gravity waves 
therefore 
curve less sharply than 
the 
circular 
orbits
of 
fluid 
particles.

The picture at wave peaks is 
more nuanced.   
To a sailor who matches 
the wave velocity, 
fluid 
appears 
slower 
at 
its 
highest 
than 
it does 
to 
sailors 
at rest 
when $1 / G < 2 r$.
In this case, $| \textbf{R} | < 
| \textbf{D} |$ since 
$| \textbf{R} | = 
| ( 1 / G ) - r | 
| \omega |$ 
while 
$| \textbf{D} | = 
r | \omega |$.
Equation~(\ref{dDdt&dRdt2}) thus predicts 
that 
the 
peaks 
of 
gravity waves 
curve more sharply than 
the 
orbits
of 
fluid particles
when 
$r G > 1 / 2$.




Even when 
it preserves 
a 
particle's 
normal 
and 
tangential
acceleration  
vectors, 
a Galilean transformation can 
thus
reconfigure a body's 
acceleration.  
As~(\ref{dDdt&dRdt2}) shows, 
the 
body's 
trajectory may sharpen 
at one extremum while broadening at 
the other.  
A trajectory 
that
exhibits 
vertical 
extrema of 
equal curvature is transformed into one which 
does not (and vice versa).

What makes this so consequential is that 
such a transformation cannot take place in 
isolation.  
Equation~(\ref{dDdt&dRdt2}) predicts that 
extrema of unequal curvature  
are 
formed by 
particles 
of unequal speed.
To 
exhibit
different speeds at peaks and troughs,
some change in speed must take 
place 
in between.
Hence, 
${\mathrm{d} | \textbf{R} |} / 
{\mathrm{d} t} \ne 0$ 
at some point(s)
between 
the  
extrema.
Since ${\mathrm{d} | \textbf{D} |} / 
{\mathrm{d} t} = 0$ everywhere,  
particles 
thus exhibit 
a form of acceleration in a frame of 
velocity \textbf{P} that 
they do not 
exhibit in a 
fixed 
frame. 
A Galilean transformation 
can 
therefore 
introduce
or remove 
a 
body's 
capacity to 
draw closer to, or pull away from, another 
body 
on the same path.





%
The total acceleration 
of Gerstner's 
particles 
is thus reconfigured 
at 
a 
trajectory's   
extrema 
and 
redistributed 
away from 
extrema.  
Table~\ref{table:props} 
highlights 
two 
pronounced 
cases: 
one 
away from the extrema, 
one 
at an extremum.  

Firstly, 
${\mathrm{d} \textbf{T}_R} / 
{\mathrm{d} t} = \textbf{0}$ 
at 
any 
inflection  
of 
a
trochoidal profile.
At 
such instants, 
particles 
exhibit 
entirely 
normal 
acceleration in a frame of 
zero velocity, but 
entirely 
tangential 
acceleration in a frame of 
velocity 
\textbf{P} 
(see~(\ref{dDdt&dRdt})).
Evidently, 
trajectories 
which 
undergo 
no change in concavity 
can 
be transformed into 
ones 
which do 
(and vice versa). 

Secondly, 
${\mathrm{d} \textbf{T}_R} / 
{\mathrm{d} t}$ does not exist at 
cusps.
Unit 
tangent 
$\textbf{T}_R$ 
does not exist at these cusps either.
As such, the right hand side of~(\ref{dDdt&dRdt}) 
breaks down at cusps, because $\textbf{T}_R$ 
and corresponding unit normal $\textbf{N}_R = 
{\mathrm{d} \textbf{T}_R} / 
{\mathrm{d} t} / \left| {\mathrm{d} \textbf{T}_R} / 
{\mathrm{d} t} \right|$ 
cannot form a basis for acceleration,  
${\mathrm{d} \textbf{R}} / 
{\mathrm{d} t}$. 
%
%
%
Thus, 
a 
regular point in a 
body's trajectory 
can be transformed 
into a singularity 
(and vice versa)
by 
moving 
between reference frames of uniform 
velocity.

It is important to note that the 
right hand side of~(\ref{dDdt&dRdt}) 
does not break down at all singular points.
Although no trochoidal node exhibits a unique 
unit tangent (e.g. see Figure~\ref{zoomin}), 
either of 
a node's 
two 
unit 
tangents can, together with its respective 
normal, form a 
basis for acceleration, 
${\mathrm{d} \textbf{R}} / 
{\mathrm{d} t}$. 

In summary, Galilean transformations 
are richer than their fundamental 
property might suggest.
Whether a body of velocity \textbf{v} 
is observed from 
a vessel 
of uniform velocity 
\textbf{f} or from 
a vessel of 
uniform velocity \textbf{g}, 
Galileo says that 
body
appears to exhibit 
an 
acceleration
equal to 
${\mathrm{d} \textbf{v}} / 
{\mathrm{d} t}$. 
That is, 
\begin{equation}
\label{galileo}
\frac{\mathrm{d} ( \textbf{v} - \textbf{f} )} 
{\mathrm{d} t}
=
\frac{\mathrm{d} ( \textbf{v} - \textbf{g} )} 
{\mathrm{d} t}
= 
\frac{\mathrm{d} \textbf{v}} 
{\mathrm{d} t} 
\end{equation}
by Galilean relativity (because 
${\mathrm{d} \textbf{f}} / 
{\mathrm{d} t}
= 
{\mathrm{d} \textbf{g}} / 
{\mathrm{d} t}
= 
\textbf{0}$).

%
%

What~(\ref{galileo}) conceals 
is the interplay between the 
components of 
${\mathrm{d} ( \textbf{v} - \textbf{f} )} 
/ 
{\mathrm{d} t}
=
{\mathrm{d} \textbf{F}} 
/ 
{\mathrm{d} t}$
and
${\mathrm{d} ( \textbf{v} - \textbf{g} )} 
/ 
{\mathrm{d} t}
=
{\mathrm{d} \textbf{G}} 
/ 
{\mathrm{d} t}$
which 
preserves~(\ref{galileo}), 
A change to any 
one of the four - 
the 
normal 
component of 
${\mathrm{d} \textbf{F}} 
/ 
{\mathrm{d} t}$,
the 
tangential
component of 
${\mathrm{d} \textbf{F}} 
/ 
{\mathrm{d} t}$,
the 
normal 
component of 
${\mathrm{d} \textbf{G}} 
/ 
{\mathrm{d} t}$,
or the 
tangential 
component of 
${\mathrm{d} \textbf{G}} 
/ 
{\mathrm{d} t}$
- must be compensated by a 
change to at least one of the other three.
By applying~(\ref{accelannotate}) to  
${\mathrm{d} \textbf{F}} / {\mathrm{d} t} =
{\mathrm{d} \textbf{G}} / {\mathrm{d} t}$, 
it is clear that,
%
\begin{equation*}
\left( 
\frac{\mathrm{d} | \textbf{F} |}{\mathrm{d} t} 
\right)^2
- 
\left( 
\frac{\mathrm{d} | \textbf{G} |}{\mathrm{d} t} 
\right)^2 
= 
| \textbf{G} |^4 \,
{{\scalebox{1.3}{$\kappa$}}}^2_G
- 
| \textbf{F} |^4 \,
{{\scalebox{1.3}{$\kappa$}}}^2_F
\end{equation*}
governs this scheme of compensation.

{\small
\bibliographystyle{unsrtnat}
\bibliography{Arxiv-Ref}

\begin{thebibliography}{27}
\providecommand{\natexlab}[1]{#1}
\providecommand{\url}[1]{\texttt{#1}}
\expandafter\ifx\csname urlstyle\endcsname\relax
  \providecommand{\doi}[1]{doi: #1}\else
  \providecommand{\doi}{doi: \begingroup \urlstyle{rm}\Url}\fi

\bibitem[von Gerstner(1804)]{Gerstner}
Franz~Joseph von Gerstner.
\newblock \emph{Theorie der wellen samt einer daraus abgeleiteten Theorie der
  Deichprofile}, pages 127--138.
\newblock Haase, 1804.

\bibitem[Henry(2008)]{Henry}
David Henry.
\newblock On {G}erstner’s water wave.
\newblock \emph{Journal of Nonlinear Mathematical Physics}, 15\penalty0 (Suppl
  2):\penalty0 87--95, 2008.

\bibitem[Froude(1861)]{Froude}
William Froude.
\newblock On the rolling of ships.
\newblock \emph{Transaction of the Institution of Naval Architects},
  2:\penalty0 180--229, 1861.

\bibitem[Rankine(1863)]{Rankine}
William John~Macquorn Rankine.
\newblock {VI}. {O}n the exact form of waves near the surface of deep water.
\newblock \emph{Philosophical transactions of the Royal society of London},
  \penalty0 (153):\penalty0 127--138, 1863.

\bibitem[Trujillo and Thurman(2017)]{Thurman}
Alan~P Trujillo and Harold~V Thurman.
\newblock \emph{Essentials of oceanography}, volume 733, page 239.
\newblock Pearson Upper Saddle River, NJ, 2017.

\bibitem[Ganot(1883)]{Ganot}
Adolphe Ganot.
\newblock \emph{Elementary Treatise on Physics Experimental and Applied for the
  Use of Colleges and Schools}, page~36.
\newblock W. Wood and Company, 1883.

\bibitem[Flick(2018)]{Flick}
Jasper Flick.
\newblock Waves, 2018.
\newblock URL \url{https://catlikecoding.com/unity/tutorials/flow/waves/}.
\newblock Accessed: 2025-10-05.

\bibitem[Thomas et~al.(2001)Thomas, Finney, Weir, and Giordano]{Thomas}
George~B Thomas, Ross~L Finney, Maurice~D Weir, and Frank~R Giordano.
\newblock \emph{Calculus}.
\newblock Boston: Addison-Wesley, 2001.

\bibitem[Phillips(2016)]{Phillips}
Emma~C Phillips.
\newblock Custom waves, 2016.
\newblock URL \url{https://www.desmos.com/calculator/8ruwco9oml}.
\newblock Accessed: 2025-10-05.

\bibitem[Danovich(2016)]{Danovich}
Mark Danovich.
\newblock Wave circles, 2016.
\newblock URL \url{https://markd87.github.io/2016/05/06/wave-circles.html}.
\newblock Accessed: 2025-10-05.

\bibitem[Raymond(2020)]{Raymond}
David~J Raymond.
\newblock Chapter 1 {W}aves in two and three dimensions, 2020.
\newblock URL \url{http://kestrel.nmt.edu/~raymond/classes/ph221/diff.pdf}.
\newblock Accessed: 2025-10-05.

\bibitem[Considine and Considine(2013)]{Considine}
Douglas~M Considine and Glenn~D Considine.
\newblock \emph{Van Nostrand’s scientific encyclopedia}, page 3138.
\newblock Springer Science \& Business Media, 2013.

\bibitem[King(2017)]{King}
George~C King.
\newblock \emph{Physics of energy sources}, pages 306--308.
\newblock John Wiley \& Sons, 2017.

\bibitem[Center(2002)]{TMD}
Thai Marine~Meteorological Center.
\newblock Chapter 1 {W}ater wave mechanics, 2002.
\newblock Accessed: 2025-10-05.

\bibitem[Toda(1989)]{Toda}
Morikazu Toda.
\newblock \emph{Nonlinear waves and solitons}, volume~5, page~15.
\newblock Springer Science \& Business Media, 1989.

\bibitem[Greenhill(1918)]{Greenhill}
Alfred~George Greenhill.
\newblock The {R}ankine trochoidal wave.
\newblock \emph{Proceedings of the Royal Society of London. Series A,
  Containing Papers of a Mathematical and Physical Character}, 94\penalty0
  (659):\penalty0 238--249, 1918.

\bibitem[Lumley(1969)]{Lumley}
John~L Lumley.
\newblock Film notes for {E}ulerian and {L}agrangian descriptions in fluid
  mechanics, 1969.
\newblock URL \url{http://web.mit.edu/hml/ncfmf/01ELDFM.pdf}.
\newblock Accessed: 2025-10-05.

\bibitem[Nelson(2008)]{Nelson}
David Nelson.
\newblock \emph{The Penguin dictionary of mathematics}.
\newblock Penguin UK, 2008.

\bibitem[Nave()]{Hyper}
Carl~Rod Nave.
\newblock Ocean waves.
\newblock URL
  \url{http://hyperphysics.phy-astr.gsu.edu/hbase/Waves/watwav2.html}.
\newblock Accessed: 2025-10-05.

\bibitem[Burgiel(2010)]{Burgiel}
Heidi Burgiel.
\newblock Cusp on the cycloid, 2010.
\newblock URL \url{http://www.mit.edu/~hlb/1802/pdf/MIT18_02SC_notes_10.pdf}.
\newblock Accessed: 2025-10-05.

\bibitem[Naik(2011)]{Naik}
Vipul Naik.
\newblock Concavity, inflections, cusps, tangents, and asymptotes, 2011.
\newblock URL
  \url{https://files.vipulnaik.com/math-152/concaveinflectioncusptangentasymptote.pdf}.
\newblock Accessed: 2025-10-05.

\bibitem[Okamoto and Shoji(2001)]{Okamoto}
Hisashi Okamoto and Mayumi Shoji.
\newblock \emph{The mathematical theory of permanent progressive water-waves},
  volume~20, page~85.
\newblock World Scientific Publishing Company, 2001.

\bibitem[Chachiyo(2025)]{Chachiyo}
Teepanis Chachiyo.
\newblock Simple and accurate complete elliptic integrals for the full range of
  modulus.
\newblock \emph{arXiv preprint arXiv:2505.17159}, 2025.

\bibitem[Sadun(2015)]{UniTexas}
Lorenzo Sadun.
\newblock Curvature and acceleration, 2015.
\newblock URL \url{https://web.ma.utexas.edu/users/m408m/Display13-4-3.shtml}.
\newblock Accessed: 2025-10-05.

\bibitem[Widnall and Peraire(2009)]{Widnall}
Sheila Widnall and Jaime Peraire.
\newblock Lecture l6 - {I}ntrinsic coordinates, 2009.
\newblock URL
  \url{https://ocw.mit.edu/courses/16-07-dynamics-fall-2009/84852a46fa77de9a750245ceb761255a_MIT16_07F09_Lec06.pdf}.
\newblock Accessed: 2025-10-05.

\bibitem[Massel(2007)]{Massel}
Stanislaw~R Massel.
\newblock \emph{Ocean waves breaking and marine aerosol fluxes}, volume~38,
  pages 11--12.
\newblock Springer Science \& Business Media, 2007.

\bibitem[Sokal(2017)]{Sokal}
Alan Sokal.
\newblock Handout \#1: Newton's first law and the principle of relativity,
  2017.
\newblock URL \url{https://www.ucl.ac.uk/~ucahad0/7302_handout_1.pdf}.
\newblock Accessed: 2025-10-05.

\end{thebibliography}
}

\end{document}